\documentclass[a4paper,11pt]{article}
\pdfoutput=1 

\usepackage{jheppub} 

\usepackage[T1]{fontenc} 
\usepackage{makecell}
\usepackage{subfigure}
\usepackage{ewfit-defs}
\usepackage{natbib}
\usepackage{graphicx}
\title{\boldmath Probing Top-quark Operators with Precision Electroweak Measurements}

\def\figureautorefname~#1\null{Fig.\,#1\null}

\def\equationautorefname~#1\null{Eq.\,(#1)\null}

\author[a,d]{Yiming Liu,}
\author[b]{Yuhao Wang,}
\author[a,d,e]{Cen Zhang,}
\author[b]{Lei Zhang,}
\author[c]{Jiayin Gu}

\affiliation[a]{Institute of High Energy Physics, Chinese Academy of Sciences, Beijing 100049, China}
\affiliation[b]{School of Physics, Nanjing University, Nanjing 210093, China}
\affiliation[c]{Department of Physics and Center for Field Theory and Particle Physics, Fudan University, Shanghai 200438, China, and \\ Key Laboratory of Nuclear Physics and Ion-beam Application (MOE), Fudan University, Shanghai 200433, China}
\affiliation[d]{School of Physical Sciences, University of Chinese Academy of Science, Beijing 100049, China}
\affiliation[e]{Center for High Energy Physics, Peking University, Beijing 100871, China}

\emailAdd{liuym@ihep.ac.cn}
\emailAdd{yuhaowang@smail.nju.edu.cn}
\emailAdd{leizhang1801@nju.edu.cn}
\emailAdd{jiayin\_gu@fudan.edu.cn}

\usepackage[dvipsnames]{xcolor}

\abstract{
In the Standard Model Effective Field Theory (SMEFT), operators involving the top quark are generally difficult to probe, 
and can generate sizable loop contributions to the electroweak precision observables, 
measured by past and future lepton colliders.  
Could the high precision of the electroweak measurements compensate the loop suppression and provide competitive reaches on these operators?  Would the inclusion of these contributions introduce too many additional parameters for a meaningful global electroweak analysis to be done?  
In this paper, we perform a detailed phenomenological study to address these two important questions.  
Focusing on eight dimension-6 operators that generate anomalous couplings between the electroweak gauge bosons and the third-generation quarks, we calculate their one loop contributions to the $e^+e^- \to f\bar{f}$ processes both on and off the Z-pole and the $e^-e^+ \to WW$ process. 
A global analysis is performed with these eight operators and the ones that contribute to the above processes at tree level, using the measurements at LEP, SLC and several low energy experiments.  
We find that, while the current electroweak precision measurements are sensitive to the one-loop effects of the top-quark operators, it is difficult to separate them from the operators that contribute at the tree level, making a global analysis rather challenging. 
Under more assumptions (for instance that the new physics contribute only to the third generation quark operators and the $S$, $T$ parameters), competitive reaches could be obtained in a global fit. 
Another important finding of our study is that the two operators that generate dipole interactions of the bottom quark have significant impacts in the Z-pole measurements and should not be omitted.  We also discuss the implication of the recently reported W-boson mass measurement at CDF to our results.  Finally, we estimate the reaches of future lepton colliders in probing the top-quark operators with precision electroweak measurements.  
}

\begin{document} 
\maketitle
\flushbottom

\section{Introduction}
\label{sec:intro}

The Standard Model (SM), despite its enormous success, is generally considered as an effective theory with a cutoff that could be as low as a few TeVs.  A lot of effort is being devoted to constructing and studying extensions of the SM that predict new particles with masses around the TeV scale. 
Direct searches for such new particles at colliders have been unfruitful so far.  
A complementary approach to the direct searches is that of indirect ones, where precise measurements of the SM processes are compared with the SM predictions, and an observed deviation from the latter is a strong indication that the process may receive virtual contributions of heavy new particles. 
A powerful model-independent framework to identify, constrain, and parametrize potential deviations with respect to the SM predictions is the Standard Model Effective Field Theory (SMEFT)\cite{Manohar:1996cq,Rothstein:2003mp,Kaplan:2005es,Burgess:2007pt,Weinberg:2009bg}. %
Assuming the electroweak symmetry breaking is linearly realized as in the SM, and the new physics scale (usually denoted as $\Lambda$) is significantly larger than the electroweak scale, the SM Lagrangian is augmented by a series of higher dimensional operators, suppressed by powers of $1/\Lambda$.  Here we focus on the effects of the leading operators (denoted as $Q_i$) that preserves baryon and lepton numbers, which are of dimension six,   
\begin{equation}
	\mathcal{L}_\mathrm{SMEFT}=\mathcal{L}_\mathrm{SM}+ \sum_{i}\frac{c_{i}Q_{i}}{\Lambda^{2}},
\end{equation}
where $\Lambda$ is the energy scale of the new physics and $c_{i}$ are the dimensionless Wilson coefficients. 
The $Q_{i}$s span the complete space of dimension-6 operators \cite{Buchmuller:1985jz, Grzadkowski:2010es, Brivio:2017vri}. If experiments detect significant deviations from SM predictions, the SMEFT could help characterize their possible origin and guide the direct searches of new physics. In the absence of any significant deviation, the SMEFT can be used to systematically constrain the scales of different BSM physics scenarios.

Many dimension-6 operators have been extensively probed by various experiments.  Among them, operators that modify the electroweak processes are stringently constrained by the precision measurements of the $Z$ and $W$ bosons at lepton colliders.  
Global analyses have been performed for these operators at both current and future colliders~\cite{Falkowski:2014tna, Efrati:2015eaa, Falkowski:2015jaa}.  While similar analyses have also been done for the Higgs and top sectors (and their combinations with the EW sector)~\cite{Durieux:2017rsg, Barklow:2017suo, Ellis:2018gqa, Durieux:2018tev, Durieux:2018ggn, Falkowski:2019hvp, deBlas:2019rxi, DeBlas:2019qco, Durieux:2019rbz, Hartland:2019bjb, Brivio:2019ius, Dawson:2020oco, Ellis:2020unq, Jung:2020uzh, Ethier:2021bye, Almeida:2021asy, Miralles:2021dyw}, some of the corresponding operators are not very well constrained due to a number of reasons.  Among them, a well-known example is the operator $(H^\dagger H)^3$, which modifies the trilinear Higgs coupling.  Even at the high luminosity LHC, its coefficient is only probed at the order-one level with the measurement of the double-Higgs process
~\cite{Cepeda:2019klc}.  Interestingly, the measurements of single-Higgs processes, where the $(H^\dagger H)^3$ operator contributes at the one-loop order, offer competitive reaches on it due to their much better measurement precision (in particular the Higgsstrahlung process at future lepton colliders)~\cite{McCullough:2013rea, Degrassi:2016wml, DiVita:2017eyz, DiVita:2017vrr}. 
Similarly, operators that generate anomalous gauge couplings of the top quark are generally less-well constrained at the tree level, and contribute to many Higgs and electroweak processes at the one-loop order~\cite{Vryonidou:2018eyv, Durieux:2018ggn, Jung:2020uzh}.  With the high precision of the electroweak measurements, it is entirely possible that the one-loop effects from the top-quark  operators are non-negligible despite the loop suppression, and should be included in the electroweak analysis.  This calls for two important questions: First, are the one-loop contributions of these top-quark operators sizable enough for the electroweak measurements to have a sensitivity that is comparable, or even better than, the ones from LHC, which probes them directly?   Second, by introducing more degrees of freedoms in the electroweak analysis, are we still able to obtain meaningful bounds in a global framework?  In other words, is it even possible to separate the effects of the top-quark operators from the tree-level electroweak ones?

In this paper, we perform a comprehensive global analysis with the current precision electroweak data in order to answer these two questions.  We focus on the effects of eight dimension-6 operators that generate anomalous couplings between the electroweak gauge bosons and the third-generation quarks.  This include the four operators that modifies the SM gauge couplings of top and bottom quarks and four operators that introduces dipole interactions. 
We include the $Z$-pole measurements at LEP and SLD, the measurements of the $e^+e^-\to WW$ process at LEP2, as well as measurements of several low energy scattering processes which are also sensitive to 4-fermion operators~\cite{Falkowski:2017pss}.  
We perform global analyses with the eight operators above and the other tree-level operators that contribute to these processes.  
For the latter, we impose the flavor universality condition in order to reduce the size of the parameter space.  
Interestingly, we also find that the tree level effects of the bottom dipole operators, while suppressed by the bottom mass, could be even larger than the one-loop contributions of the other operators, and these tree level effects need  to be included for consistency.  
As a first step towards a more complete global analysis, we do not consider the effects of any top-quark-related four-fermion operators in our study.  These operators introduce additional degrees of freedoms, and in many cases are notoriously difficult to separate from other top operators~\cite{Durieux:2018tev, Durieux:2019rbz, Miralles:2021dyw}.  Recent studies also show that they have significant impacts on Higgs processes~\cite{Alasfar:2022zyr}.

The rest of this paper is organized as follows:  In \autoref{sec:Theory}, we lay out the theoretical framework of our study, including the operators we consider, the corresponding tree-level and one-loop contributions to the electroweak processes, and details of the Monte Carlo simulation we used to obtain some of the results.  In \autoref{sec:Experiment}, we provide a detailed description of the experimental inputs used in our analysis.  Our results are presented in \autoref{sec:FitResult}. 
We consider both the general framework and a more restrictive ``semi-universal'' scenario.  Results with the new CDF-II W mass measurement~\cite{CDF:2022hxs} are  presented, and its implication is also discussed. 
Our projections for the future lepton colliders are provided in \autoref{sec:FutureCollider}.  Finally, we draw conclusion in \autoref{sec:Conclusion}.  
More details of our analysis and additional results are presented in \autoref{sec:Appendix}.

\section{Theory Framework}
\label{sec:Theory}

\subsection{The SMEFT framework}
\label{sec:smeft}

We work in a global SMEFT framework and consider all dimension-6 operators that contribute to the current electroweak measurements (listed in \autoref{sec:Experiment}).  
Our standpoint is that the third generation quarks are special.  Operators involving the third generation quarks are thus separated from the rest of the 
dimension-6 operators, and their one-loop contributions are considered, in addition to possible tree-level ones.  For the other operators, we consider only the tree-level contributions\footnote{Strictly speaking, this partial-loop-contribution framework we implement is not basis-independent, as the coefficients of the third-generation-quark operators could become a linear combination involving other operator coefficients under a basis transformation.  
However, we find the impacts of such effects 
numerically irrelevant as long as the other operators already contribute at the tree level.   
}, and impose a  $U(2)_{u}$ $\bigotimes$ $U(2)_{d}$ $\bigotimes$ $U(2)_{q}$ $\bigotimes$$U(3)_ {l}$ $\bigotimes$ $U(3)_e$ flavor symmetry. 
This setup allows us to investigate the impacts of the third-generation-quark operators while maintaining a relatively small parameter space.  As mentioned earlier, four-fermion operators involving the top quark are not included in our study.   
We leave a more general analysis with the four-fermion top-quark operators and less 
flavor assumptions to future studies.  

\begin{table}[htbp]
    \centering
    \resizebox{\textwidth}{!}{
    \begin{tabular}{|l|l|l|}
        \hline
        $\psi ^ { 2 } \varphi ^ { 3 }$&$X^{3}$&$\varphi ^ { 4 } D ^ { 2 }$\\
        \hline
        $Q_{u\varphi}^{ i j}=(\varphi^\dag \varphi)(\bar q_i u_j
        \widetilde\varphi)$&$Q_W=\epsilon^{IJK} W_\mu^{I\nu} W_\nu^{J\rho}
        W_\rho^{K\mu}$&$Q_{\varphi D}=\left(\varphi^\dag D^\mu\varphi\right)^\star
        \left(\varphi^\dag D_\mu\varphi\right)$ \\
        $Q_{d\varphi}^{ i j}=(\varphi^\dag \varphi)(\bar q_i d_j \varphi)$& & \\
        \hline
        $\psi ^ { 2 } \varphi ^ { 2 } D$&$\psi ^ { 2 } X \varphi$&$X ^ { 2 } \varphi ^
        { 2 }$\\
        \hline
        $Q_{\varphi l}^{ij(1)}=\left(
        \varphi^{\dagger}i\stackrel{\leftrightarrow}{D}_{\mu}\varphi\right)\left(
        \overline{l}_{i}\gamma^{\mu}l_{j} \right)$&$Q_{uW}^{i j}=(\bar
        q_{i}\sigma^{\mu\nu}u_{j})\tau^{I}\widetilde{\varphi}W_{\mu\nu}^{I}$&$Q_{\varphi
            WB}=\varphi^{\dag}\tau^{I} \varphi W^{I}_{\mu\nu} B^{\mu\nu}$ \\
        $Q_{\varphi l}^{ij(3)}=\left(
        \varphi^{\dagger}i\stackrel{\leftrightarrow}{D}_{\mu}^{I}\varphi \right)\left(
        \overline {l}_{i}\tau^{I}\gamma^{\mu}l_{j} \right)$&$Q_{uB}^{ij}=(\bar
        q_{i}\sigma^{\mu\nu}u_{j}) \widetilde\varphi B_{\mu\nu}$& \\
        $Q_{\varphi e}^{ i j}=\left( \varphi ^ { \dagger } i \stackrel {
            \leftrightarrow } { D } _ { \mu } \varphi \right)(\bar e_i \gamma^\mu
        e_j)$&$Q_{dW}^{ i j}=(\bar q_i \sigma^{\mu\nu} d_j) \tau^I \varphi\,
        W_{\mu\nu}^I$&\\
        $Q_{\varphi q}^{ i j(1)}=\left( \varphi ^ { \dagger } i \stackrel {
            \leftrightarrow } { D } _ { \mu } \varphi \right)(\bar q_i \gamma^\mu
        q_j)$&$Q_{dB }^{i j}=(\bar q_i \sigma^{\mu\nu} d_j) \varphi B_{\mu\nu}$&\\
        $Q_{\varphi q}^{ i j(3)}=\left( \varphi ^ { \dagger } i \stackrel {
            \leftrightarrow } { D } _ { \mu } ^ { I } \varphi \right)(\bar q_i \tau^I
        \gamma^\mu q_j)$&&\\
        $Q_{\varphi u}^{ i j}=\left( \varphi ^ { \dagger } i \stackrel {
            \leftrightarrow } { D } _ { \mu } \varphi \right)(\bar u_i \gamma^\mu u_j)$&&\\
        $Q_{\varphi d}^{ i j}=\left( \varphi ^ { \dagger } i \stackrel {
            \leftrightarrow } { D } _ { \mu } \varphi \right)(\bar d_i \gamma^\mu d_j)$&&\\
        $Q_{\varphi u d}^{ i j}=i(\widetilde\varphi^\dag D_\mu \varphi)(\bar u_i
        \gamma^\mu d_j)$&&\\
        \hline
        $( \overline { L } L ) ( \overline { L } L )$&$( \overline { R } R ) (
        \overline { R } R )$&$( \overline { L } L ) ( \overline { R } R )$\\
        $Q_{ll}^{ p r st }=(\bar l_p \gamma_\mu l_r)(\bar l_s \gamma^\mu
        l_t)$&$Q_{ee}^{ p r s t}=(\bar e_p \gamma_\mu e_r)(\bar e_s \gamma^\mu
        e_t)$&$Q_{le}^{ p r s t}=(\bar l_p \gamma_\mu l_r)(\bar e_s \gamma^\mu e_t)$ \\
        $Q_{lq }^{p r s t(1)}=(\bar l_p \gamma_\mu l_r)(\bar q_s \gamma^\mu
        q_t)$&$Q_{eu}^{ p r s t}=(\bar e_p \gamma_\mu e_r)(\bar u_s \gamma^\mu
        u_t)$&$Q_{lu}^{ p r s t}=(\bar l_p \gamma_\mu l_r)(\bar u_s \gamma^\mu u_t)$\\
        $Q_{lq }^{p r s t(3)}=(\bar l_p \gamma_\mu \tau^I l_r)(\bar q_s \gamma^\mu
        \tau^I q_t)$&$Q_{ed}^{ p r s t}=(\bar e_p \gamma_\mu e_r)(\bar d_s\gamma^\mu
        d_t)$&$Q_{ld }^{p r s t}=(\bar l_p \gamma_\mu l_r)(\bar d_s \gamma^\mu d_t)$ \\
        &&$Q_{qe }^{p r s t}=(\bar q_p \gamma_\mu q_r)(\bar e_s \gamma^\mu e_t)$ \\
        \hline
    \end{tabular}}
    \caption{
    The operators in the Warsaw basis~\cite{Grzadkowski:2010es} that are used in our study. The indices $i$, $j$ and $p$, $r$, $s$, $t$ label the fermion generation.} 
    \label{table_operators}
\end{table}

Under the above assumptions, the dimension-6 operators involved in our study are summarized in \autoref{table_operators}, where the Warsaw basis are used~\cite{Grzadkowski:2010es}.  These operators are divided into two classes.  The first class involves the third generation quarks, which are
\begin{equation}
Q^{(1)}_{\varphi Q} \,,~~~~ Q^{(3)}_{\varphi Q} \,,~~~~ Q_{\varphi t} \,,~~~~ Q_{\varphi b} \,,~~~~ 
Q_{tW} \,,~~~~ Q_{tB} \,,~~~~ Q_{bW} \,,~~~~ Q_{bB} \,,  \label{eq:Q1}
\end{equation}
where, instead of writing down the flavor indices $ij=33$, we have made the replacement $q\to Q$, $u\to t$, $d\to b$. Here, $Q,~t,~b$ denotes the $SU(2)$ doublet $(t,\,b)_L$ and the singlets $t_R$, $b_R$, respectively. 
The corresponding Wilson coefficients, $c_i/\Lambda^2$, follow the same notation.  The four operators $Q^{(1)}_{\varphi Q}$, $Q^{(3)}_{\varphi Q}$, $Q_{\varphi t}$ and $Q_{\varphi b}$ modify the SM gauge couplings between the 3rd generation quarks and the $W$, $Z$ bosons.  There is another operator, $Q_{\varphi tb} = i(\widetilde\varphi^\dag D_\mu \varphi)(\bar t_i\gamma^\mu b_j)$, which generates a right-handed $Wtb$ coupling.  It contributes only to the $W$-boson self energy, and this contribution is strongly suppressed by the bottom mass. %
For this reason, we do not include it in our analysis.  The rest four operators, $Q_{tW}$, $Q_{tB}$, $Q_{bW}$ and $Q_{bB}$, generate dipole interactions between the 3rd generation quarks and the gauge bosons.  Due to their different helicities, a fermion mass insertion is needed to generate an interference term with the SM amplitude.  For the top quark, this contribution could certainly be sizable and should be included in the analysis.  The impacts of the bottom-quark dipole operators are more subtle --- while their contributions to the $Zb\bar{b}$ vertex are suppressed by the small (but non-negligible) bottom mass, these are tree level contributions and turn out be comparable or even larger than the one-loop contributions of top operators. 
As such, they are also included in our analysis.  %
For all eight operators, 
 their leading contributions are included into all the observables in a consistent way. That is, if an operator already contributes to a process at the tree level (such as the bottom-quark operators to $e^+e^-\to b\bar{b}$), we shall only consider its tree-level contribution; otherwise, its one-loop contribution is included.  

The second class contain all other operators that contribute to the $e^+e^- \to f \bar{f}$ and $e^+e^- \to WW$ processes at the tree level,  %
including the ones that modify the propagators of $Z$ and $W$ bosons, their couplings to fermions (excluding the 3rd generation quarks) and the triple gauge couplings, 
\begin{align}
&Q^{(1)}_{\varphi q} \,,~~~~ Q^{(3)}_{\varphi q} \,,~~~~ Q_{\varphi u} \,,~~~~ Q_{\varphi d} \,,~~~~ Q^{(1)}_{\varphi l} \,,~~~~ Q^{(3)}_{\varphi l} \,,~~~~ Q_{\varphi e} \,, \nonumber\\
& Q'_{ll} \,,~~~~ Q_{\varphi D}  \,,~~~~ Q_{\varphi WB} \,,~~~~ Q_W\,, 
\end{align}
as well as the 4-fermion operators that directly contribute to the $e^+e^-\to f\bar{f}$ processes and several low energy scattering processes, 
\begin{align}
&Q_{qe}  \,,~~~~ Q_{eu}  \,,~~~~ Q_{ed}  \,,~~~~ Q^{(1)}_{lq}  \,,~~~~  Q^{(3)}_{lq}  \,,~~~~  Q_{lu}  \,,~~~~ Q_{ld}  \,, \nonumber\\
& Q_{ll}   \,,~~~~ Q_{ee}  \,,~~~~  Q_{le} \,,
\end{align}
where 
the flavor indices are omitted due to the flavor symmetries we imposed.  
Note that we have distinguished $Q'_{ll} \equiv Q^{1221}_{ll}$ (which contributes to the $\mu$ decay) and $Q_{ll}  \equiv Q^{11ii}_{ll}$ (which contributes to $e^+e^- \to l^+ l^-$) even though they are not independent in the flavor universal case.  
This is because many new physics models that contribute to the latter (such as a flavor-diagonal $Z'$ boson) do not contribute to the former.  
In addition, we include the 4-fermion operators in the $e^+e^- \to b\bar{b}$ process in the second class and only consider their tree-level effects. 
They are
\begin{equation}
Q^{(1)}_{lQ}, \,~~~ Q^{(3)}_{lQ}, \,~~~ Q_{lb} \,,~~~~  Q_{eQ} \,,~~~~ Q_{eb} \,,
\end{equation}
where we have again made the replacement $q\to Q$, $u\to t$, $d\to b$ instead of writing down the 3rd generation flavor indices. 
For convenience, we also use the following combinations of operators instead of the original ones for the 3rd generation quarks,
\begin{align}
Q^{(+)}_{\varphi Q} \equiv Q^{(1)}_{\varphi Q} + Q^{(3)}_{\varphi Q} \,, & \hspace{1cm}  Q^{(+)}_{l Q} \equiv Q^{(1)}_{l Q} + Q^{(3)}_{l Q} \,,  \nonumber \\
Q^{(-)}_{\varphi Q} \equiv Q^{(1)}_{\varphi Q} - Q^{(3)}_{\varphi Q} \,, & \hspace{1cm}  Q^{(-)}_{l Q} \equiv Q^{(1)}_{l Q} - Q^{(3)}_{l Q} \,,
\end{align}
and their Wilson coefficients follow the same labels.  $ Q^{(-)}_{l Q}$ contributes only to a contact $e^+e^- t\bar{t}$ interaction and is not considered in our analysis.  In total, 33 Wilson coefficients are included in our analysis, and we consider their leading contributions for each observable, which is at the one-loop level if there is no tree level contribution.

It is well known that, in the Warsaw basis, the operator coefficients that contribute to the $Z$-pole observables at the tree level exhibit flat directions, such that a global SMEFT fit with only the $Z$-pole observables (and the $W$ mass measurement) could not be closed.  These flat directions are lifted by the measurements of the $e^+e^-\to WW$ process~(see {\it e.g.} \cite{Falkowski:2014tna}).  
However, since the measurement precision of $e^+e^-\to WW$ at LEP is significantly worse than the $Z$-pole ones,\footnote{This is partially due to the fact that the measurements of the $W$-decay angles are not available in the final report~\cite{ALEPH:2013dgf}, so we could only use the production polar angle distribution in our analysis.  See \autoref{sec:Experiment} for more details.} 
there still remain large correlations among many operator coefficients.  This could obscure the impacts of 
the 3rd-generation-quark operators that are the focus of this study.  
To resolve this issue, in most parts of our analysis we use a slight different basis, obtained by replacing operators $Q_{\varphi D}$, $Q_{\varphi WB}$ with 
\begin{equation}
Q_{D \varphi W} =  i D_{\mu} \phi^{\dagger} \sigma_{a} D_{v} \phi W^{a \mu v}\,, \hspace{1.5cm} Q_{D \varphi B} =  i D_{\mu} \phi^{\dagger} D_{v} \phi B^{\mu v} \,,
\end{equation}
which do not contribute to the $Z$-pole observables.  The translation to the Warsaw basis is given by
\begin{align}
Q_{D \varphi B} 
\to& ~ -\frac{g^{\prime}}{4}  Q_{\varphi B}+\frac{g^{\prime}}{2} \sum_{\psi} Y_{\psi} Q_{\varphi \psi}^{(1)}+\frac{g^{\prime}}{4} Q_{\varphi \square}+g^{\prime} Q_{\varphi D}-\frac{g}{4} Q_{\varphi W B} \,, \nonumber \\
Q_{D \varphi W} 
\to & ~~ \frac{g}{4} \sum_{F} Q_{\varphi F}^{(3)}+\frac{g}{4}\left(3 Q_{\varphi \square}+8 \lambda_{\phi} Q_{\varphi}-4 \mu_{\phi}^{2}\left(\phi^{\dagger} \phi\right)^{2}\right)+ \nonumber \\
&+\frac{g}{2}\left(y_{i j}^{e}\left(Q_{e \varphi}\right)_{i j}+y_{i j}^{d}\left(Q_{d \varphi}\right)_{i j}+y_{i j}^{u}\left(Q_{u \varphi}\right)_{i j}+\text { h.c. }\right) \nonumber \\
&-\frac{g^{\prime}}{4} Q_{\varphi W B}-\frac{g}{4} Q_{\varphi W} \,.
\end{align}

%

We use the following input parameters in our analysis~\cite{Zyla:2020zbs}:  
\begin{gather}
    \alpha=\frac{1}{127.9},\,~~~~ m_{Z}=91.1876\,\GeV, \,~~~~ G_{F}=1.166379\times 10^{-5}\,\GeV, \notag \\
    m_{b}=4.7\,\GeV,\,~~~~ m_{t}=172.5\,\GeV \,.
    \label{eq:input}
\end{gather} 

\subsection{Loop contributions}
\label{sec:loop}

\begin{figure}[t]
    \centering
    \includegraphics[width=0.4\linewidth]{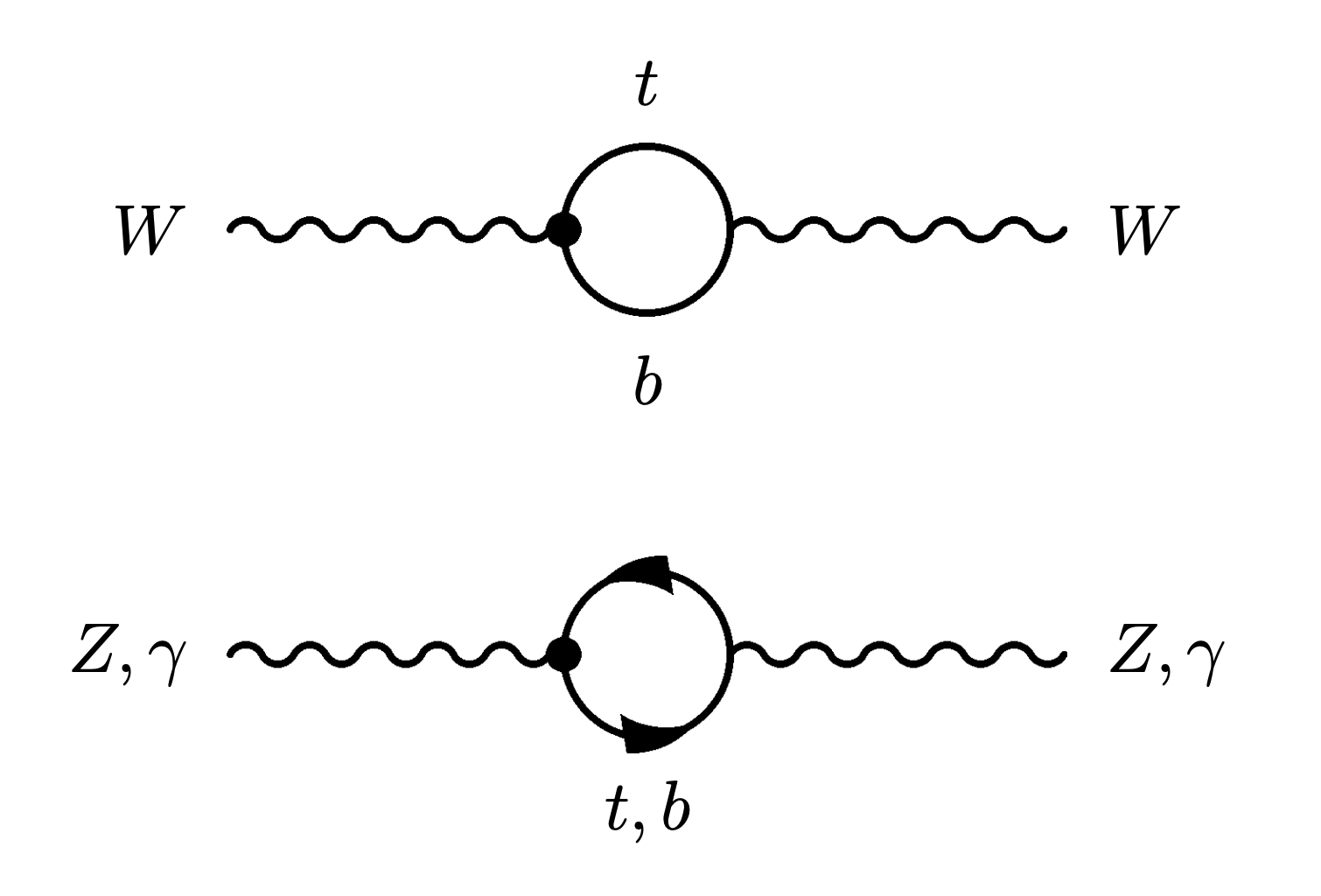}
    \caption{The gauge boson self-energy correction caused by the operators in \autoref{eq:Q1}.
        The black dots are the dimension-6 vertex.}
    \label{loop_feynman}
\end{figure}

The tree level contributions of the SMEFT dimension-6 operators to the $e^+e^- \to f\bar{f}$ and  $e^+e^- \to WW$ processes have been well studied in the past~\cite{Falkowski:2014tna, Efrati:2015eaa, Falkowski:2015jaa, Hagiwara:1993ck, Han:2004az, Berthier:2015oma, Berthier:2015gja}, 
and we follow these references in our analysis. %
Here we focus on the one-loop contributions of the 3rd-generation-quark operators in \autoref{eq:Q1}. %
First, they generate universal contributions to the gauge boson self-energies, as shown in \autoref{loop_feynman}.  They have both direct and indirect effects.  
The indirect ones are the contributions to the measurements of the three input parameters, $\alpha$, $m^2_Z$ and $G_F$.  Same as the tree level dimension-6 contributions, they change the ``inferred SM values'' of these parameters.  Their effects can be parameterized as
\begin{align}
\alpha&=\alpha _{O}\left(1+\Pi^{'}_{\gamma\gamma}(0)\right) \,, \nonumber\\
m^{2}_{Z}&=m^{2} _{ZO}+\Pi_{ZZ}(m^{2}_{Z}) \,, \nonumber\\
G_{F}&=G _{FO}\left(1-\frac{\Pi_{WW}(0)}{m_{W}^{2}}\right)\,,  \label{eq:O1}
\end{align}
where $\alpha$, $m^{2}_{Z}$ and $G_{F}$ take the values in \autoref{eq:input}, $\alpha _{O}$, $m^{2} _{ZO}$ and $G _{FO}$ are the renormalized SM parameters that enter the calculations of EW observables, and $\Pi^{(i)}_{XY}(q^2)$ parameterizes the one-loop corrections to the gauge boson self-energies~\cite{Peskin:1991sw}, for which we only consider the contributions from the operators in \autoref{eq:Q1}.  Their definitions are shown in \autoref{def_renormalized_par}.  A renormalized $s^2_W \equiv \sin^2\theta_W$ can be defined as
\begin{align}
\notag
s^{2} _{WO}&=\frac{1}{2}\Bigg(1-\sqrt{1-\frac{4\pi\alpha _{O}}{\sqrt{2}G _{FO}m _{ZO}^{2}}}\Bigg)\\
&=s_{W}^{2}\Bigg(1-\frac{c^{2}_{W}}{c^{2}_{W}-s^{2}_{W}}\Big(\Pi^{'}_{\gamma\gamma}(0)+\frac{\Pi_{WW}(0)}{m^{2}_{W}}-\frac{\Pi_{ZZ}(m^{2}_{Z})}{m^{2}_{Z}}\Big)\Bigg),  \label{eq:O2}
\end{align}
where $s_{W}^{2}=\frac{1}{2}\Big(1-\sqrt{1-\frac{4\pi\alpha}{\sqrt{2}G_{F}m_{Z}^{2}}}\Big)$.  The $W$, $Z$ self-energies also directly enter the observables.  The contributions to $e^+e^- \to f\bar{f}$ as well as the $W$, $Z$ decay rates can be characterized by the tree-level neutral and charged current interactions.  At tree level, they are given by (assuming no tree-level contributions from dimension-6 operators) 
\begin{align}
M_{NC}&=e^{2}_{O}\frac{QQ^{'}}{q^{2}}+\frac{e^{2}_{O}}{s_{WO}^{2}c_{WO}^{2}}(I_{3}-s_{WO}^{2}Q)\frac{1}{q^{2}-m_{ZO}^{2}}(I^{'}_{3}-s_{WO}^{2}Q^{'}) \,, \nonumber\\
M_{CC}&=\frac{e^{2}_{O}}{2s_{WO}^{2}}I_{+}\frac{1}{q^{2}-m_{WO}^{2}}I_{-} \,, \label{eq:nccco}
\end{align}
where $I^{(')}_{3}$, $Q^{(')}$  are the $SU(2)$ and electric charges of the external fermions, $I_{+}, I_{-}$ are the isospin-raising and isospin-lowering matrices, and $\alpha _{O}$, $m^{2} _{ZO}$, $G _{FO}$, $s^{2} _{WO}$ are the renormalized SM parameters in \autoref{eq:O1} and \autoref{eq:O2}.  
With the direct contributions from the $W$, $Z$ self-energies, \autoref{eq:nccco} is modified into
\begin{align}
    M^{1}_{NC}&=e_{*}^{2}\frac{QQ^{'}}{q^{2}}+\frac{e_{*}^{2}}{s_{W^{*}}^{2}c_{W^{*}}^{2}}(I_{3}-s_{W^{*}}^{2}Q)\frac{Z_{Z^{*}}}{q^{2}-m_{Z^{*}}^{2}}(I^{'}_{3}-s_{W^{*}}^{2}Q^{'})  \,, \nonumber\\
    M^{1}_{CC}&=\frac{e_{*}^{2}}{2s_{W^{*}}^{2}}I_{+}\frac{Z_{W^{*}}}{q^{2}-m_{W^{*}}^{2}}I_{-} \,,
\end{align}
where $m^2_{W*} (q^2)$, $m^2_{Z*} (q^2)$, $Z_{W*} (q^2)$, $Z_{Z*} (q^2)$, $s^2_{W*} (q^2)$, $e^2_* (q^2)$ generally depend on $q^2$ of the propagators, and their expressions are also listed in \autoref{def_renormalized_par}\cite{Zhang:2012cd}.  
\begin{figure}[t]
    \centering
 \includegraphics[width=0.4\linewidth]{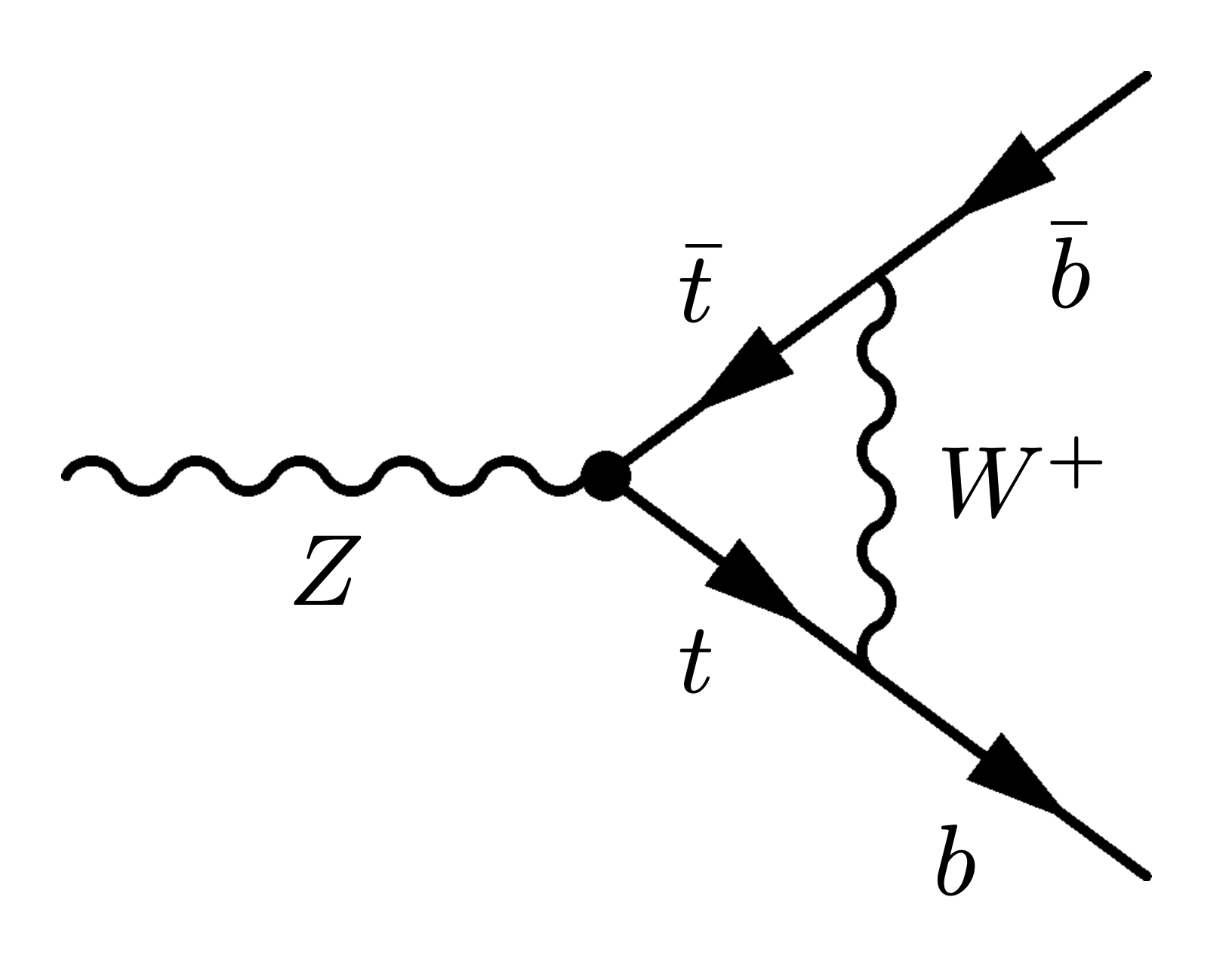}
  \caption{Top loop diagram modification to the $bb$ production.}
  \label{toploop_bb}  
\end{figure}
%
In additional to the $W$, $Z$ self-energies, the 3rd-generation-quark operators also directly modifies the $Zb\bar{b}$ vertex as shown in \autoref{toploop_bb}, thus generating non-universal effects.  
%
Finally, they also contribute to the three gauge boson vertex, which enters $e^+e^-\to WW$.  The calculation of these effects are rather complicated, for which we rely on Monte Carlo integration with MadGraph5\_aMC@NLO~\cite{Alwall:2014hca}.  
The method of rewighting \cite{van2011oneloop} is used to generate weighted events. The SM input parameters used in Monte Carlo are same as \autoref{eq:input}. %
We also checked that the statistical uncertainties due to simulation sample size are sufficiently small and can be neglected.  
The numerical results of our calculation are be found in github repository\footnote{https://github.com/yuhao-wang-nju/electroweak-data-collection}. 

The one-loop contributions to EW observables from dimension-6 operators are also recently studied in Refs.~\cite{Dawson:2019clf, Dawson:2022bxd}.  In Ref.~\cite{Dawson:2019clf}, the one-loop order QCD and electroweak corrections to the $Z$ and $W$ pole observables are computed under the flavor universality assumption.  In contrast, our study focuses on the 3rd generation quarks. By also computing the one-loop contributions of the first two generation quarks, and then imposing the flavor universality assumption, we managed to reproduce the results in Ref.~\cite{Dawson:2019clf} on the contributions of $O_{\varphi t}$, $O_{tW}$ and $O_{tB}$ to the $Z\rightarrow bb$ decay width. For $O_{\varphi Q}^{(-)}$, our result turns out to be about 4 times larger than the one in Ref.~\cite{Dawson:2019clf}, 
which is probably due to the different choice of cutoff energy. 
Ref.~\cite{Dawson:2022bxd} studied the one-loop corrections of the flavor non-universal 4-fermion interactions, and is thus complementary to our study.

\section{Experiment}
\label{sec:Experiment}
The data used in this study is mainly from the LEP experiment. Some data of the low energy experiment like CHARM \cite{allaby1987precise}, CDHS \cite{blondel1990electroweak}, CCFR \cite{mcfarland1998precision}, NuTeV \cite{zeller2002precise}, APV \cite{wood1997measurement}, QWEAK \cite{gericke2009q} and PVDIS \cite{wang2014measurement} is also included. It can be divided into the following category:

\begin{itemize}
    \item Precision Electroweak Measurement: We use the $Z$-pole measurement from LEP experiment, $W$ mass measurement taken from combined results of PDG Group and branching ratios (BR) information from LEP.
    \item $e^+ e^- \to f \bar{f}$: The measurements of electron collision to quark pair and lepton pair from the LEP experiment are collected. Some observables measured by TRISTAN at 58G$eV$ are also included.
    \item $e^+ e^- \to W^+ W^- \to 4f$: The total cross-section and differential cross-section measurements of $e^+ e^- \to W^+ W^- \to 4f$ from the LEP experiment are included in this study.
    \item Low Energy Measurement: The neutral-current parameters for $\nu$-hadron and $\nu-e$ processes measured by deep inelastic scattering (DIS) experiments including CDHS and CHARM in CERN, CCFR and NuTeV in FermiLab. The parameters for electron scattering is measured by the Cs atomic parity violation (APV) experiment, electron-deuteron DIS (eDIS) experiment PVDIS and Qweak that measure the weak charge of proton. Additionally, the measurement of the cross-section of trident production ($\nu_\mu \gamma^* \rightarrow \nu_\mu \mu^+ \mu^-$) over its SM prediction by CHARM and CDHS is included.
\end{itemize}

For the data mentioned above, we take the center values, uncertainties as well as the correlations. The correlations between different categories and experiments are taken as zero. For those that can not find the information about their correlations, we make the consumption that they are not correlated with each other. Table\,\ref{observables_table} shows the varieties of observables included in this study and the references.
\begin{table}[htbp]
    \centering
    \resizebox{\textwidth}{!}{
    \begin{tabular}{|c|c|c|c|}
        \hline
        & Experiment & Observables  & Reference \\ \hline
        Low Energy & \makecell[c]{CHARM/CDHS/\\CCFR/NuTeV/\\APV/QWEAK/\\PVDIS} & Effective Couplings & \makecell{\cite{Zyla:2020zbs},\cite{geiregat1990first}}\\ \hline
        $Z$-pole & LEP/SLC & \makecell[c]{Total decay width $\Gamma_Z$\\ \hline Hadronic cross-section $\sigma_{had}$\\ \hline Ratio of decay width $R_f$\\ \hline Forward-Backward Asymmetry $A_{FB}^f$\\ \hline Polarized Asymmetry $A_f$} & \makecell{\cite{efrati2015electroweak}}\\ \hline
        $W$-pole & \makecell{LHC/Tevatron/\\LEP/SLC} & \makecell{Total decay width $\Gamma_W$ \\ \hline $W$ branching ratios $Br(W\rightarrow l v_l)$ \\ \hline Mass of $W$ Boson $M_W$} & \makecell{\cite{efrati2015electroweak},\cite{Zyla:2020zbs}}\\ \hline
        $ee\rightarrow qq$ & LEP/TRISTAN & \makecell{Hadronic cross-section $\sigma_{had}$ \\ \hline Ratio of cross-section $R_f$ \\ \hline Forward-Backward Asymmetry for $b$/$c$ $A_{FB}^f$} & \makecell{\cite{aleph2013electroweak},\cite{Alcaraz:2006mx},\cite{inoue2000measurement}}\\ \hline 
        $ee \rightarrow ll$ & LEP & \makecell{cross-section $\sigma_f$ \\ \hline Forward-Backward Asymmetry $A_{FB}^f$ \\ \hline Differential cross-section $\frac{d\sigma_f}{dcos\theta}$} & \makecell{\cite{aleph2013electroweak}}\\ \hline
        $ee\rightarrow WW$ & LEP & \makecell{cross-section $\sigma_{WW}$\\ \hline Differential cross-section $\frac{d\sigma_{WW}}{dcos\theta}$} & \makecell{\cite{aleph2013electroweak}}\\ \hline
    \end{tabular}}
    \caption{Observables used in this study}
    \label{observables_table}
\end{table}

Table~\ref{tree_level_op} lists the type-I operators that have impact on the above-mentioned observables. Besides these operators, five operators ( $Q_{\varphi WB}$, $Q^{ijji}_{ll}$, $Q_{\varphi D}$, $Q^{11(3)}_{\varphi l}$, $Q^{22(3)}_{\varphi l}$) have impact on all the obersvables as they will affect the electroweak parameters: $g$, $g^{'}$, $v$. In the following part, details about some categories involved with experiments will be discussed. These details also affect the generation of Monte Carlo events. Additionally, the uncertainties of SM calculation and experimental measurement will be discussed.

\subsection{\texorpdfstring{$Z$, $W$ pole observables and low energy couplings}{Z, W pole observables and low energy couplings}}
The theoretical prediction of $Z$ pole observables and $W$ pole observables are taken from the Table 10.5 of \cite{Zyla:2020zbs} and the Table 2 of \cite{Efrati:2015eaa}. The theoretical errors of $Z$, $W$ pole observables are ignored and low energy couplings, except for $m_{W}$ (the theoretical prediction error of $m_{W}$ is 0.004GeV \cite{Zyla:2020zbs,Baak:2014ora}).\subsection{Fermion-Pair Production}
\par
\begin{figure}[htbp]
    \centering
        \subfigure[s-channel ee$\rightarrow$ff]{
        \begin{minipage}[t]{0.45\linewidth}
        \centering
        \includegraphics[height=1in]{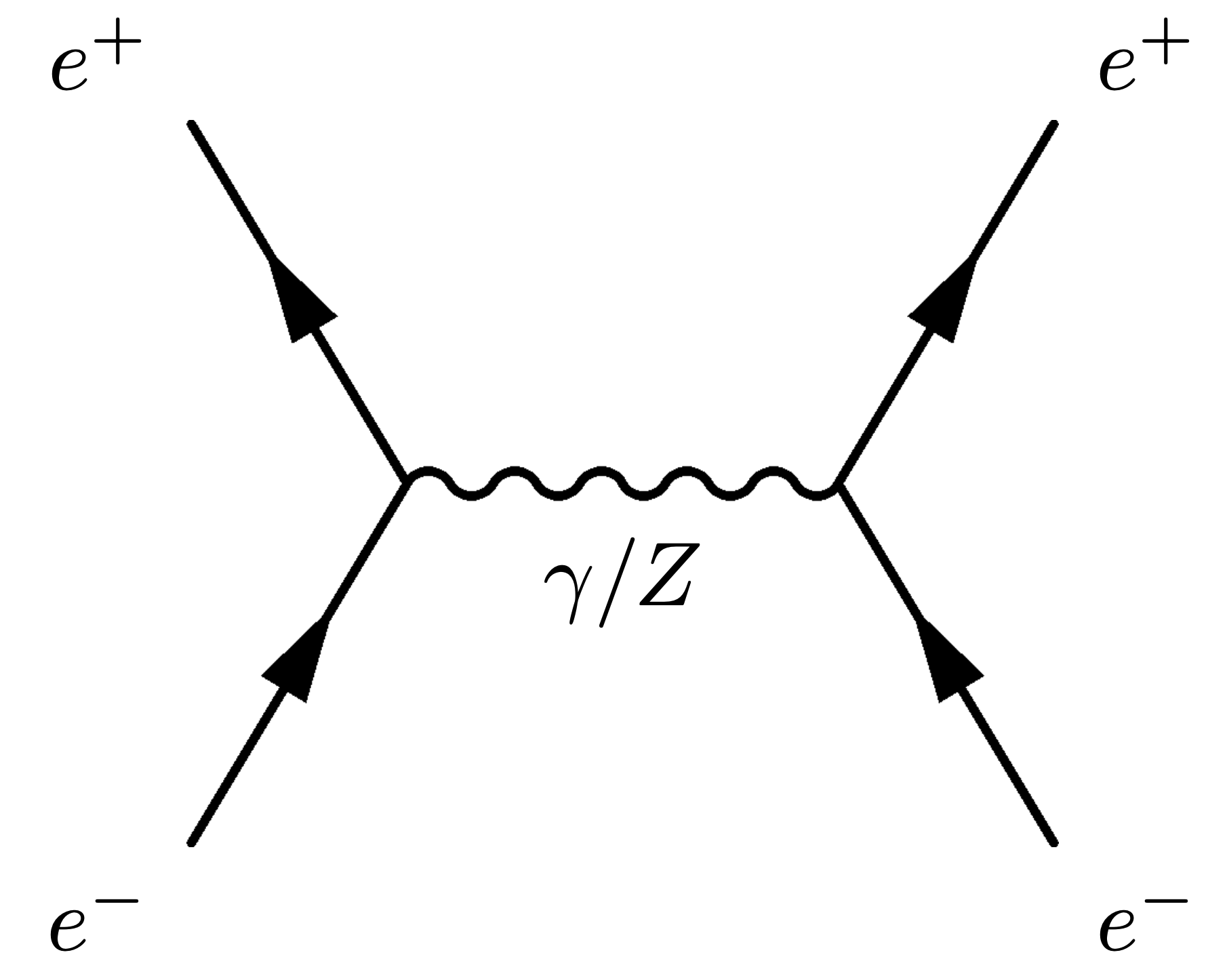}
        \end{minipage}
        \label{eeffannil}
    }
    \subfigure[t-channel ee$\rightarrow$ee]{
        \begin{minipage}[t]{0.45\linewidth}
        \centering
        \includegraphics[height=1in]{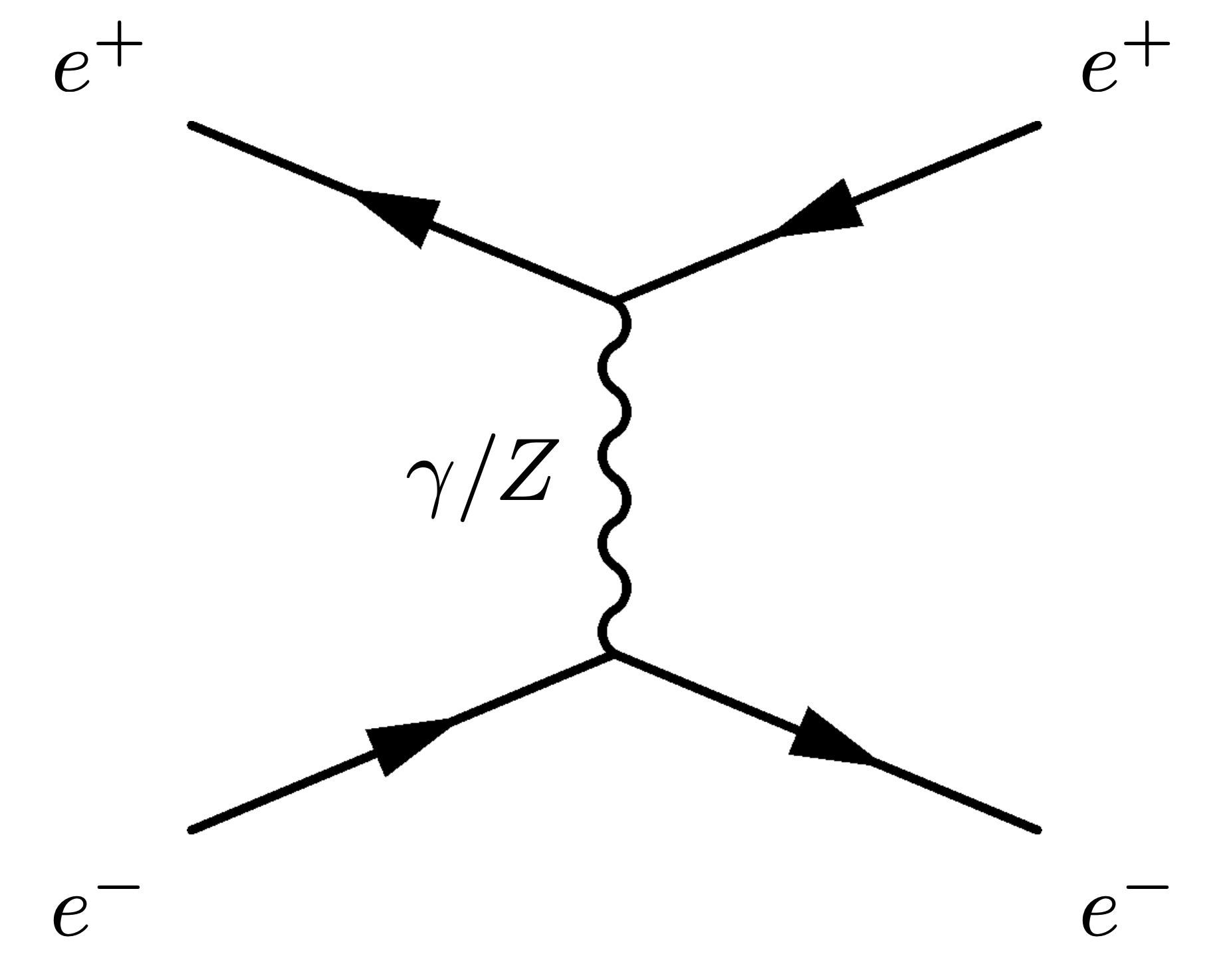}
        \end{minipage}
        \label{eeexchange}
    }
    \caption{Feynman diagrams for process $e^+ e^- \to f \bar{f}$ at Born level}
\end{figure}

The main feynman diagram of $e^+ e^- \rightarrow f \bar{f}$ is the s-channel photon and $Z$ annihilation (Figure\,\ref{eeffannil}). However, there are additional t-channel photon and $Z$ exchange in $e^+ e^- \rightarrow e^+ e^-$ process (Figure\,\ref{eeexchange}), which have a quite small contribution at $Z$-pole. We make the assumption of zero mass of electron and muon in our model. As a result, there would be only a negligible difference between the contribution of new operators to observables of $e^+ e^- \rightarrow e^+ e^-$ and $e^+ e^- \rightarrow \mu^+ \mu^-$ process, and we will take the result of $e^+ e^- \rightarrow \mu^+ \mu^-$ as that of $e^+ e^- \rightarrow e^+ e^-$ at $Z$-pole. At other energies, both the t-channel and s-channel production of $e^+ e^- \rightarrow e^+ e^-$ are considered.

For quark pair production, the decay of quark is not considered. Additional observables in low energy region for $b$ and $c$ quark production are included, which would help us get a better constraint on the 3rd-generation-quark operators.

For the fermion-pair production, especially at the energies above $Z$ resonance, the QED radiative corrections are very large. This is caused by initial-state radiation of photons, which will lower the center-of-mass energy, $\sqrt{s}$, to the value $\sqrt{s'}$. In the LEP experiment, the measurement is done with the events which have small amount of initial state radiation, \textit{i}.\textit{e}., large $\sqrt{s/s'}$. So in 
the Monte Carlo integration, initial-state radiation is not considered.  

\begin{itemize}
\item In the processes of $e^{+}e^{-}\to f\bar{f}$($f=\mu, \tau, q$), the experimental values of cross-sections(in pb) and forward-backward asymmetries are reported in Table 3.4 of \cite{aleph2013electroweak}. The theoretical prediction uncertainties of $\sigma(q\bar{q}), \sigma(\mu^{+}\mu^{-})$,  $\sigma(\tau^{+}\tau^{-})$ and leptonic forward-backward asymmetries are $0.26$\%, $0.4$\%, $0.4$\% and $0.4$\% respectively \cite{aleph2013electroweak,Kobel:2000aw}. Their theoretical prediction uncertainties can be neglected because their experimental uncertainties are at least four times larger. 
\item The experimental and theoretical value of the $e^{+}e^{-}\to\mu^{+}\mu^{-}$, $e^{+}e^{-}\to\tau^{+}\tau^{-}$ differential cross-section are reported in Table 3.8 and 3.9 of \cite{aleph2013electroweak}. In the processes of $e^{+}e^{-}\to\mu^{+}\mu^{-}$, $e^{+}e^{-}\to\tau^{+}\tau^{-}$, we assume the theoretical uncertainties of differential distribution are 0.4\%, which can be neglected compared with experimental ones.
\item The experimental and theoretical value of the  $e^{+}e^{-}\to e^{+}e^{-}$ differential cross-section are reported in Table 3.11 and 3.12 of \cite{aleph2013electroweak}. The theoretical uncertainties of the large-angle Bhabha scattering for $\sqrt{s}$ ranging from 189 G$eV$ to 207 G$eV$ are around 0.5\% \cite{Caffo:1997yy,Kobel:2000aw} . Theoretical uncertainties of differential distribution are taken as 0.5\% as well. 
\item In the process of $e^{+}e^{-}\to q\bar{q}$ at $\sqrt{s}=58$ G$eV$, the Table 8 in \cite{Inoue:2000hc} show the experimental value for $R_{b}$ and $R_{c}$, which are the bottom and charm quark pair production cross-section ratio to the total hadronic cross-section. In \cite{Berthier_2016}, their theoretical uncertainties are estimated as 1\% and can be ignored.
\item In the process of $e^{+}e^{-}\to q\bar{q}$ at $\sqrt{s}$ ranging from 189 G$eV$ to 207 G$eV$, the Table 8.9, 8.10 in \cite{Alcaraz:2006mx} show the experimental values for $R_{b}$ and $R_{c}$ without the theoretical uncertainties. We assume the theoretical uncertainties are negligible. 
\end{itemize}

\subsection{Boson-Pair Production}
\begin{figure}[htbp]
    \centering
    \includegraphics[width=0.8\linewidth]{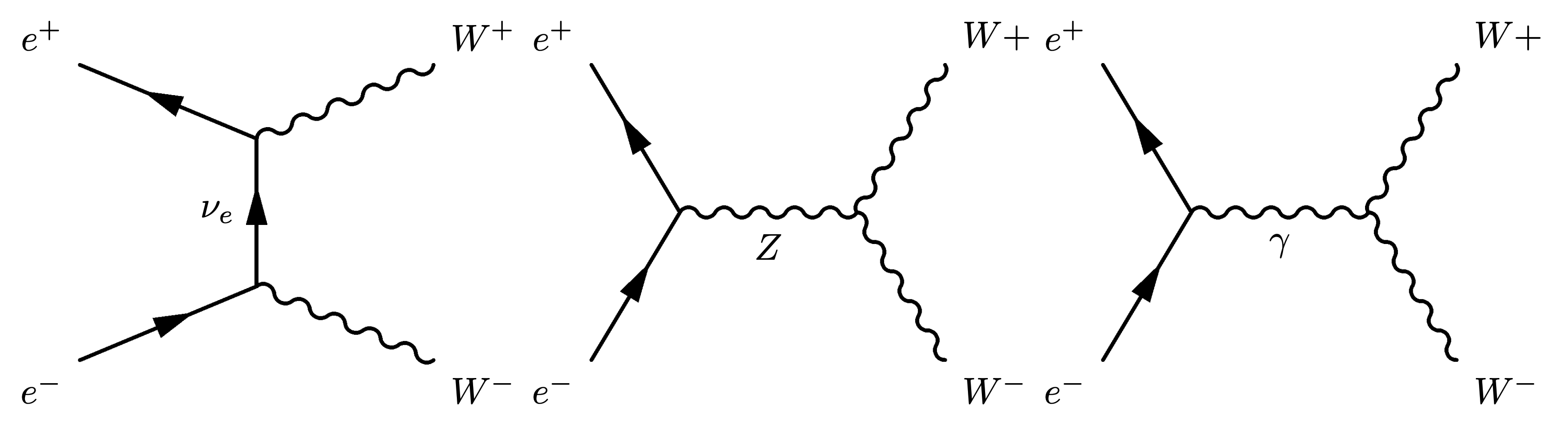}
    \caption{Main feynman diagrams for $e^+ e^- \rightarrow W^+ W^-$ at Born level}
    \label{cc03}
\end{figure}

The boson-pair production is also very important in LEP experiment. In our study, $e^+ e^- \rightarrow W^+ W^-$ process is included. There are three main feynman diagrams of this process, shown in figure\ref{cc03}, which are named as CC03 (Charged Current).

The $W$ boson will decay to a quark-antiquark pair or a lepton-neutrino pair. The branch ratio of hadronic decay is about 67\% and that of leptonic decay is around 33\%. The $WW$ events are classified into fully hadronic, semi-leponic and fully leptonic events. In the Monte Carlo simulation of the total cross-section of $WW$ production, we ignore the decay of $W$ boson and only produce the events that take $W$ pair as the final state for convenience.

\begin{figure}[htbp]
    \centering
    \subfigure[\cpDC]{
        \begin{minipage}[t]{0.45\linewidth}
            \centering
            \includegraphics[width=\linewidth]{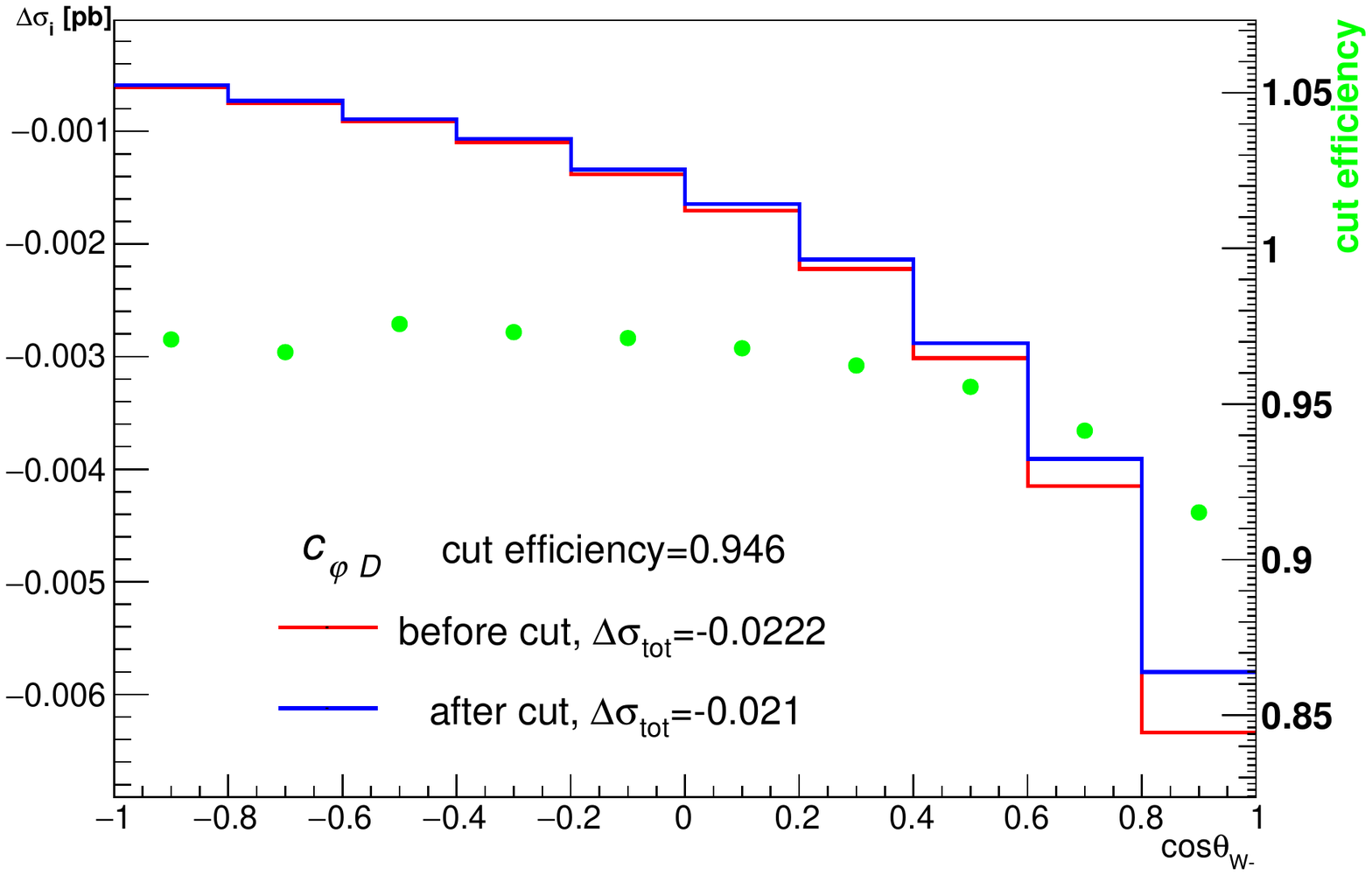}   
        \end{minipage}
    }
    \subfigure[\cpe]{
        \begin{minipage}[t]{0.45\linewidth}
            \centering
            \includegraphics[width=\linewidth]{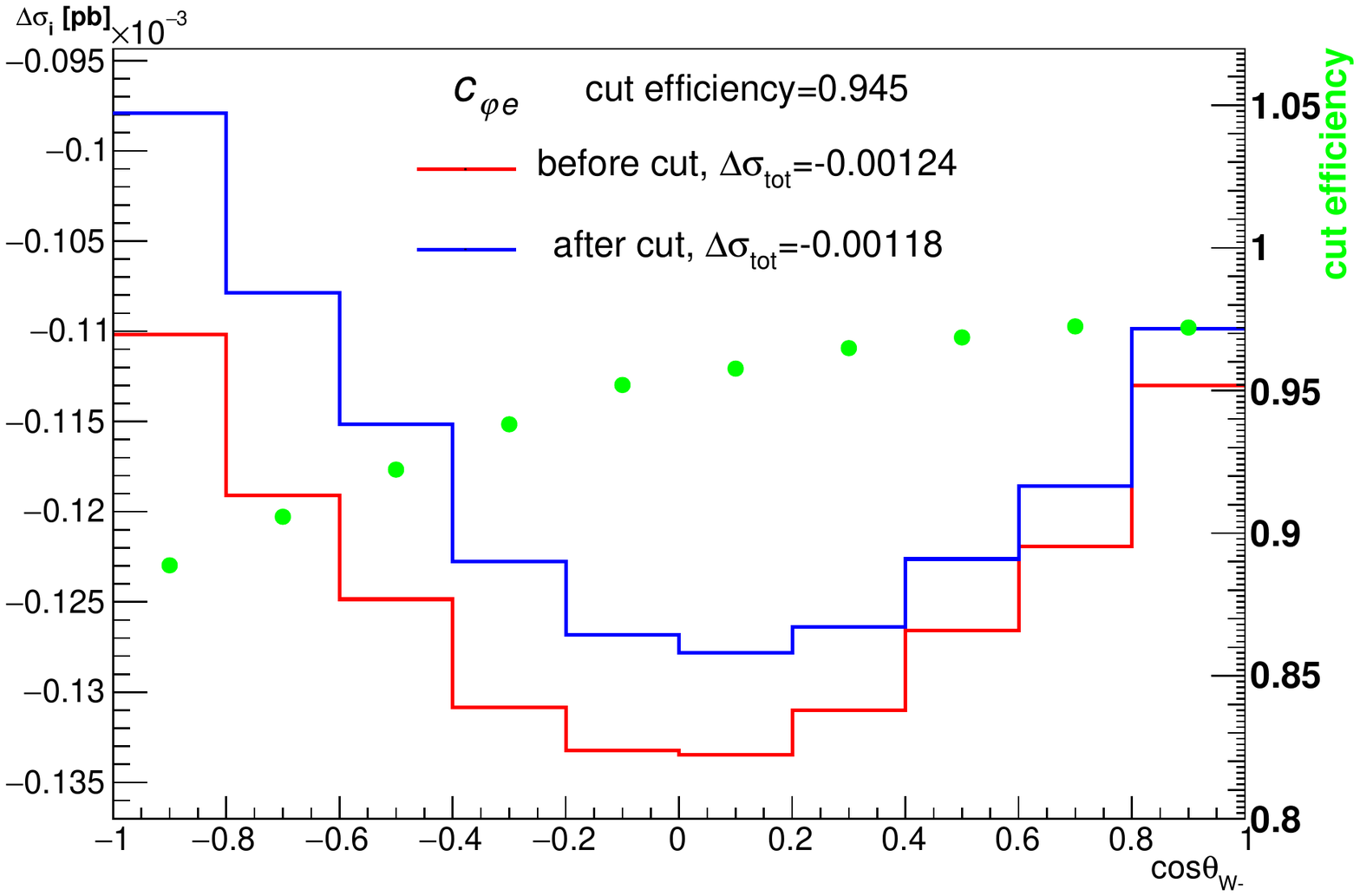}   
        \end{minipage}
    }
    \caption {20 degree angular cut's impact on the operator \cpDC\,and \cpe\, at 183 G$e$V. The red lines show the differential cross-section of the operators before the angular cut and the blue line show the differential cross-section after the cut. The overall cut efficiency is given by the total cross-section after the cut divided by the one before cut.}
    \label{angularcut_cpDC}
\end{figure}

The angular distribution of $WW$ events ($d\sigma_{WW}/dcos\theta_{W^-}$) is also included in this study, where $\theta_{W^-}$ is the polar angle between $W^-$ and the $e^-$ beam direction. In the differential cross-section measurement, the decay of $WW$ is selected as $q\bar{q}e\nu$ and $q\bar{q}\mu\nu$ for the reason that these two kinds of leptonic decay provides the charge tag and a clean background compared to $\tau$ channel. There is a constraint in the experiment that the charged lepton should be $20^\circ$ away from the beam, $|\theta_l|>20^\circ$. This angular cut brings about a 93\% cut efficiency in the experiment,\cite{aleph2013electroweak} which matches our Monte Carlo simulation with SM model. This angular requirement corresponds to the experimental acceptance of the four LEP experiments and also greatly reduces the difference between the full 4$f$ cross-section and the CC03 cross-section by reducing the contribution of t-channel diagrams in the $q\bar{q}e\nu$ final state (As stated in the LEP result, the difference of 4$f$ and CC03 cross-section reduced from 24.0\% to about 3.5\% with the angular cut). However, its effect is no longer the same to the contribution from new physics. According to our simulation, the angular cut efficiency of some operators is shown in the figure\,\ref{angularcut_cpDC}. The cut efficiency of the operators which contribute to the differential cross-section at different energies can be found in the appendix. As a result, the decay of $W$ can not be ignored in the simulation of $W$ angular distribution. Taking the four fermions as the final state, the process of $e^+ e^- \rightarrow W^+ W^- \rightarrow q \bar{q} l \nu$ would be a subset of the process $e^+ e^- \rightarrow q \bar{q} l \nu$. If we simulate the process $e^+ e^- \rightarrow W^+ W^- \rightarrow q \bar{q} l \nu$ and allow the $W$ boson be off-shell, the gauge invariance would be broken. On the other side, simulating the process $e^+ e^- \rightarrow q \bar{q} l \nu$ would bring additional feynman diagram. Finally, we decided to simulate the process $e^+ e^- \rightarrow W^+ W^- \rightarrow q \bar{q} l \nu$ using MadSpin\cite{Artoisenet_2013} to put $W$ on shell decay in order to guarantee gauge invariance.

In \cite{Grunewald:2000ju,Jadach:2001mp}, the theoretical uncertainties of $W^{-}$ angular distribution are around $0.5$\% when $\sqrt{s}$ ranges from 180 G$e$V to 210 G$e$V. Since the experimental precision of $W^{-}$ differential angular cross-section are seven times larger than theoretical ones at least, we can neglect the theoretical prediction uncertainties of $W^{-}$ differential angular cross-section. Although the theoretical uncertainties of total cross-section of $W$ pair production are also around $0.5$\% for $\sqrt{s}$ ranging from 180 G$e$V to 210 G$e$V, they are non-negligible because the experimental uncertainties are comparable with the theoretical ones.
\section{Fit Result}
\label{sec:FitResult}
\subsection{Methodology}
The SMEFT wilson coefficients are estimated with the method of least squares under the assumption of Gaussian errors. The $\chi^2$ function is constructed as: 
\begin{equation}
    \chi^2=(\vec{y}-A\vec{c})^TV^{-1}(\vec{y}-A\vec{c}) \,,
\end{equation}
where $\vec{y}$ is the vector of the difference between the experimental result and SM prediction of observables, $A$ is the contribution matrix($A_{ij}$ is the i-th operator's contribution to j-th observables), $\vec{c}$ is the vector of the Wilson coefficient and $V$ is the covariance matrix of observables.

Let $\nabla\chi^2=0$, we can get the Least Square Estimator $\hat{\vec{c}}\;$ that minimize $\chi^2$ and its covariance matrix $U$:
\begin{equation}
    \hat{\vec{c}}=(A^TV^{-1}A)^{-1}A^TV^{-1}\vec{y}=B\vec{y}
\end{equation}
\begin{equation}
    U=BVB^T=(A^TV^{-1}A)^{-1}
\end{equation}
The element of covariance matrix, $U_{ij}$, stands for the covariance of the estimator $\hat{c_i}$ and $\hat{c_j}$. The one-sigma bound of $c_i$ can be obtained from the diagonal element of this matrix.  


\subsection{Global analysis}
\label{sec:ga}

We apply the aforementioned fit strategy to obtain the constraints of the EFT operator coefficients. Two kinds of bounds are shown: one is the marginalized bound, derived allowing all the operator coefficients to float.  When conducting this kind of fit, the correlations between the coefficients 
also contain useful information.  
The other is the individual bound, which is obtained by considering only one operator coefficient at a time while fixing all others to zero. In the following results, all the bounds 
are given in 68\% confidence level (CL).  
The new physics scale $\Lambda$ in Wilson coefficients $\frac{c_i}{\Lambda^2}$ is fixed to be 1\,TeV throughout this section.  

\begin{figure}[htbp]
    \centering
    \includegraphics[width=\linewidth]{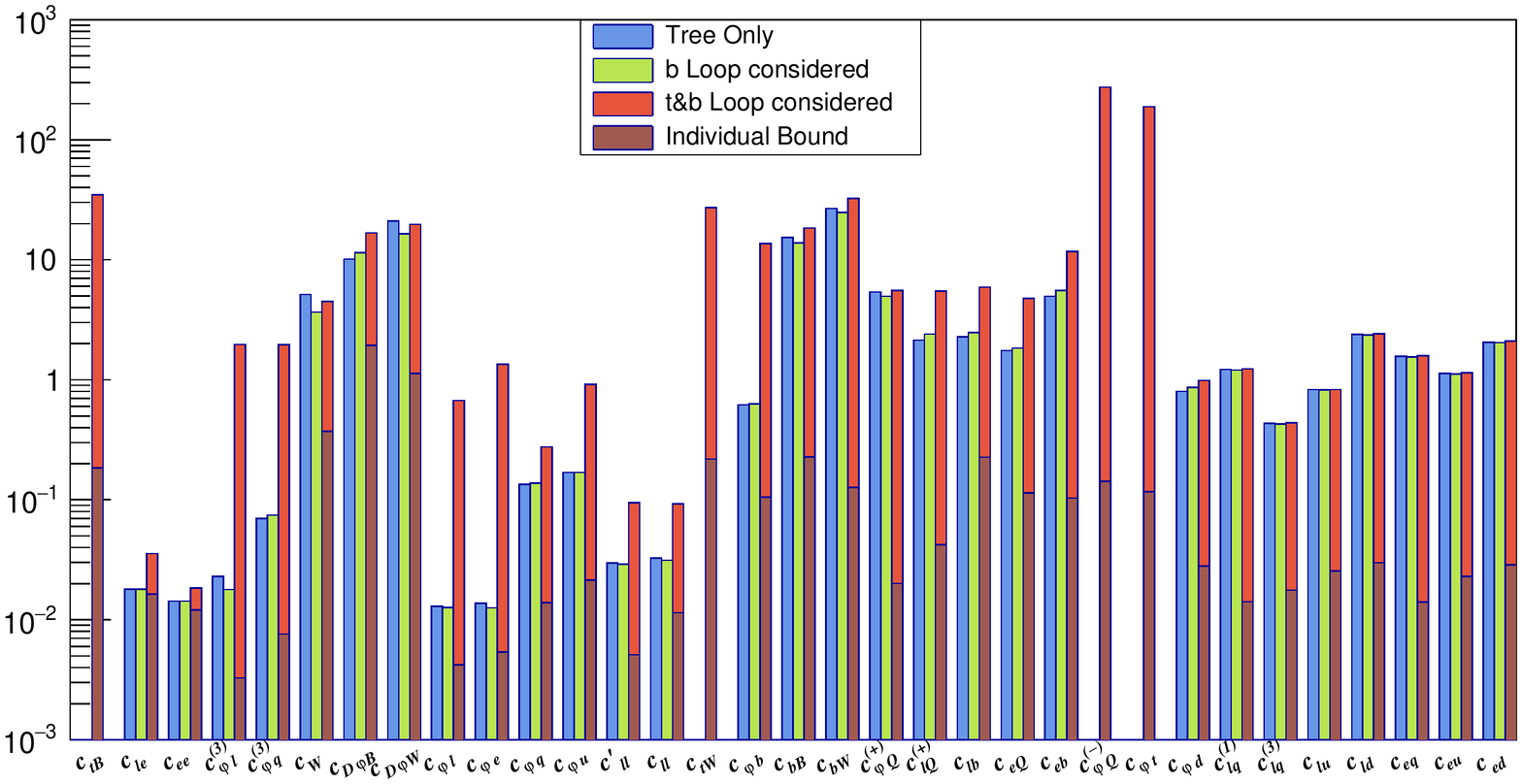}
    \caption{The 68\%CL bounds of the coefficients of operators in modified Warsaw basis for three different scenarios. Blue: Tree-level contribution only. Green: Tree-level contribution and the loop contributions from the bottom-quark operators. Red: Tree-level contribution and the loop contributions of all third-generation-quark operators listed in  \autoref{sec:Theory}, the dark red bar is the individual bound in this scenario. See \autoref{numerical_fit_result} for the best-fitted central values. The value of $\Lambda$ is set to 1\,TeV for all coefficients.}
    \label{barplot}
\end{figure}

\begin{figure}[htbp]
    \centering
    \includegraphics[width=\linewidth]{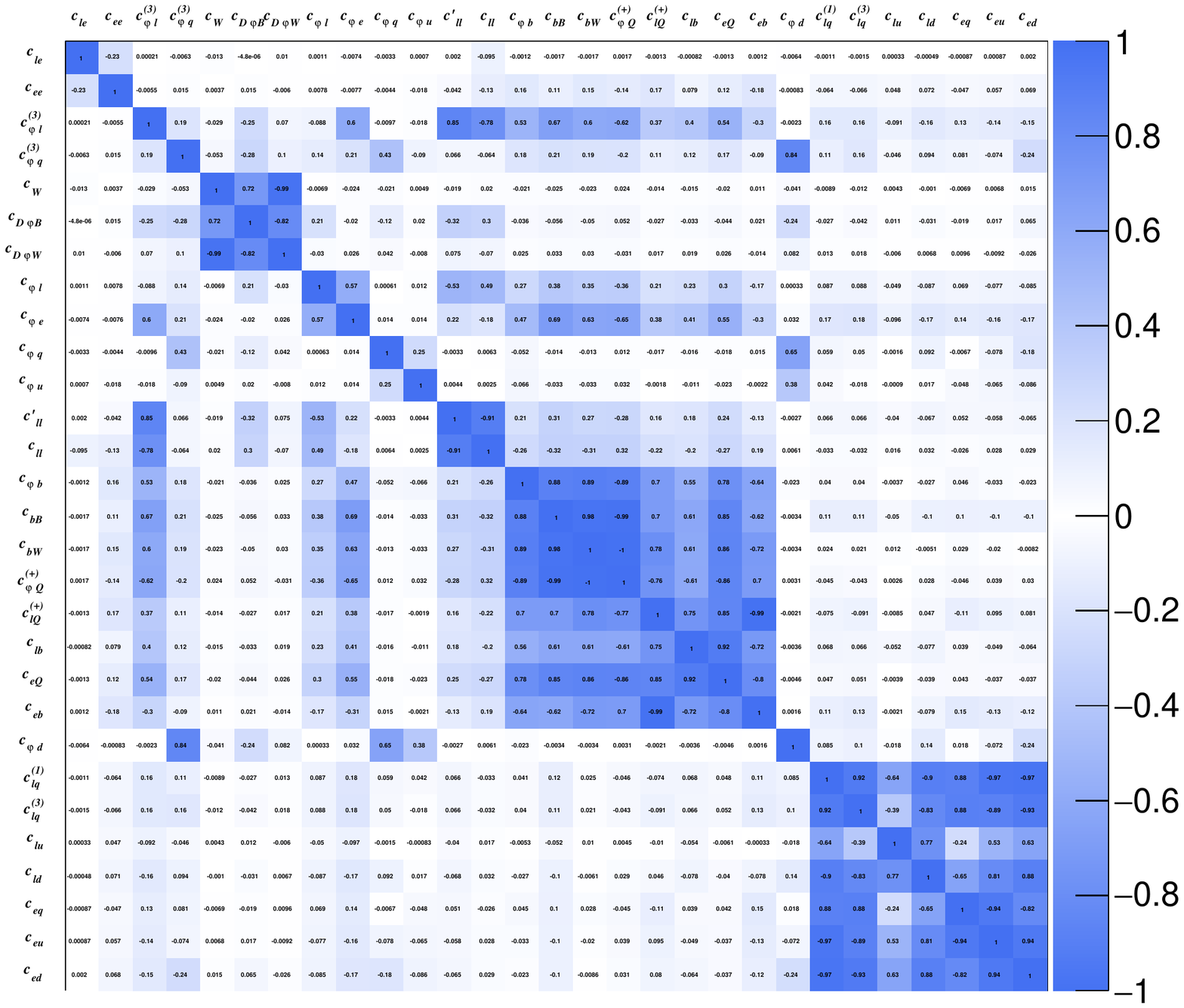}
    \caption{Correlation matrix between operators' Wilson coefficients in modified Warsaw basis. Only the tree-level contribution of these operators are considered.}
    \label{34tree_newbasis_corr}
\end{figure}

\begin{figure}[htbp]
    \centering
    \includegraphics[width=\linewidth]{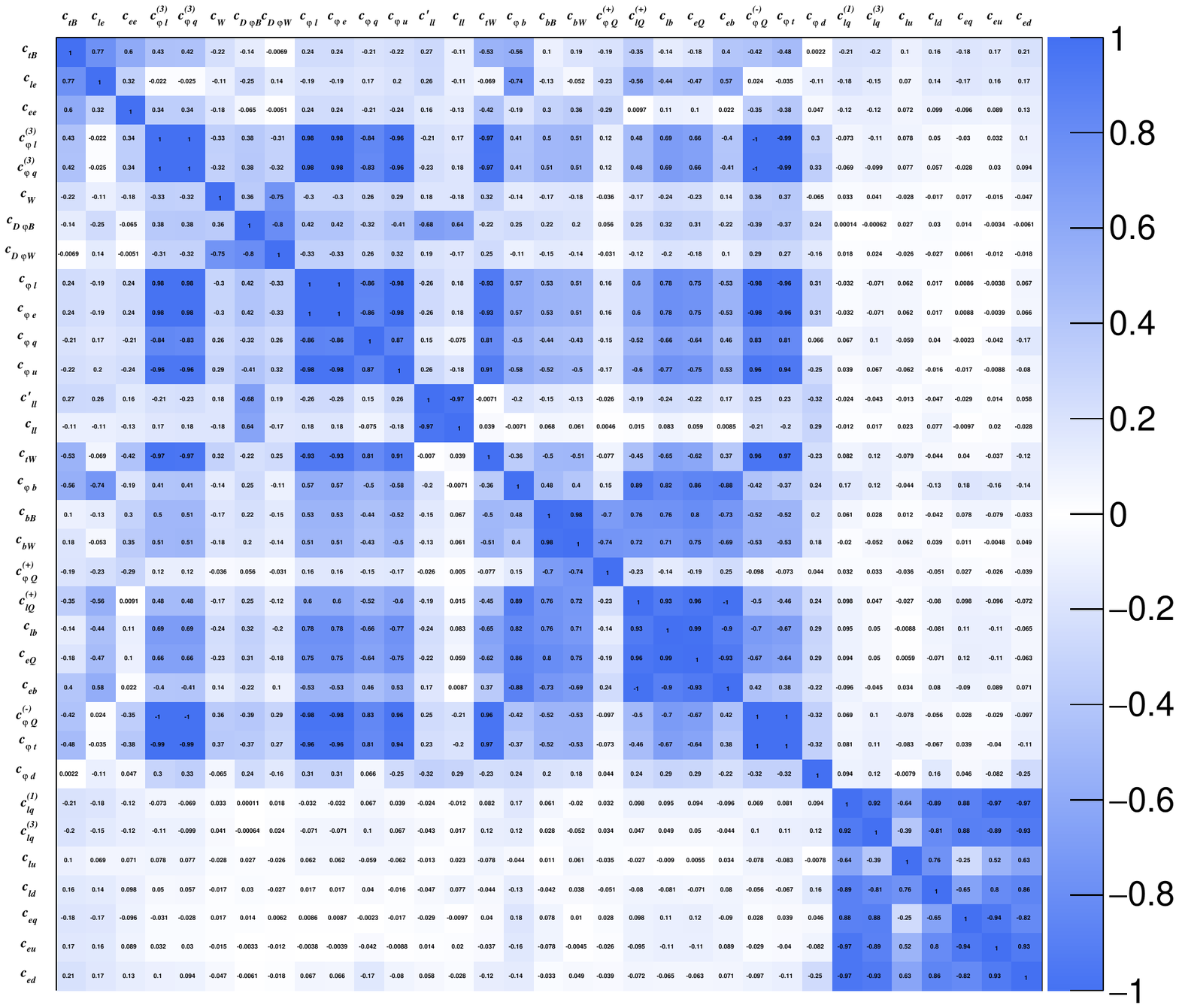}
    \caption{Correlation matrix between operators' Wilson coefficients in modified Warsaw basis. The top and bottom loop contribution are considered in addition to the tree-level contribution.}
    \label{34loop_newbasis_corr}
\end{figure}

We consider a total number of 33 operators listed in \autoref{sec:Theory} which are constrained by the measurements in \autoref{sec:Experiment}.  As mentioned in \autoref{sec:Theory}, we trade $\frac{c_{\varphi D}}{\Lambda^{2}}Q_{\varphi D}$ and $\frac{c_{\varphi WB}}{\Lambda^{2}}Q_{\varphi WB}$ in the Warsaw basis for $\frac{c_{D\varphi B}}{\Lambda^{2}}iD^{\mu}\varphi^{\dagger}D^{\nu}\varphi B_{\mu\nu}$ and $\frac{c_{D\varphi W}}{\Lambda^{2}}iD^{\mu}\varphi^{\dagger}\sigma_{a}\varphi W^{a}_{\mu\nu}$ in order to disentangle some large correlations, while the rest of the operators in the Warsaw basis remain unchanged.  Our main results are presented in \autoref{barplot} in terms of the one-sigma (corresponding to a 68\% confidence level (CL)) bounds of the Wilson coefficients. To understand the impacts of the 3rd generation quark operators, we also consider two additional scenarios for comparison: one considers only tree-level contributions from dimension-6 operators, and the other also includes the loop contributions of the bottom-quark operators %
({\it i.e.}~excluding $Q^{(-)}_{\varphi Q}$,  $Q_{\varphi t}$, $Q_{tW}$, $Q_{tB}$).  The correlations between the coefficients are shown in \autoref{34tree_newbasis_corr} for the tree-level-only scenario and \autoref{34loop_newbasis_corr} for the full scenario. The numerical results can be found in \autoref{numerical_fit_result} and the fisher information matrix for the full scenario is shown in \autoref{allfisher}.

Overall, our results demonstrates the relevance of the 3rd generation quarks in the electroweak measurements.  
The coefficients of the four top operators $Q^{(-)}_{\varphi Q}$,  $Q_{\varphi t}$, $Q_{tW}$, $Q_{tB}$ are constrained to be in the range of  $0.1$ to $1$ 
for the individual fit, despite the fact that their contributions only enter at the one-loop order.  Assuming an order-one coupling, this corresponds to a new physics scale of a few TeVs.  
After comparing these results with the ones in \cite{Ethier:2021bye} which uses the LHC data to probe these four top operators, it is found that the electroweak data have greater constraining power on $Q^{(-)}_{\varphi Q}$ and $Q_{\varphi t}$, whose advantage is estimated to be order of 10. While for $Q_{tW}$ and $Q_{tB}$, the LHC data is more powerful. Their 95\% CL individual bounds are listed in \autoref{tab:comparison_with_LHC} for comparison.
	\begin{table}[h]
		\centering
		\begin{tabular}{|c|cccc|}
			\hline
			  &
			$c_{\varphi t}$&$c_{\varphi Q}^{(-)}$&
			$c_{tW}$ & $c_{tB}$
			\\\hline
		Electroweak&
			0.233 & 0.286 & 0.438& 0.36 \\
		 LHC data&2.275 &  1.22 & 0.06 & 0.145
			\\\hline
		\end{tabular}
		\caption{95\% CL individual bounds of top quark operators' coefficients using different sets of observables.The electroweak results are the ones in this study while the LHC refers to the results in \cite{Ethier:2021bye}.
		}	
		\label{tab:comparison_with_LHC}
	\end{table}

In a global fit, however, the marginalized bounds of the top-quark operator coefficients become much looser, and some of them have possibly exceeded the range of EFT validity.  This is not surprising, as the introduction of additional degrees of freedom tends to bring additional flat directions, making the fit difficult to converge.  This can also be verified by comparing the correlations in \autoref{34tree_newbasis_corr} and \autoref{34loop_newbasis_corr}, as the increase of correlation is rather visible in the latter.  The inclusion of top operators also significantly degrade the reaches of some other operator coefficients.  Among them, the most notable ones are the leptonic operators: $c_{l e}$, $c_{e e}$, $c^{(3)}_{\varphi l}$, $c_{\varphi l}$, $c_{\varphi e}$ (as well as $c^{(3)}_{\varphi q}$), which were previously very well constrained and are thus most sensitive to additional degrees of freedom.  
Note that, $c_{l e}$, $c_{e e}$ contribute directly to the 4-lepton processes, and their contributions can be distinguished from the ones with a gauge boson propagator  
by measuring the processes at several different energies.
On the other hand, when only the loop contributions of the bottom-quark operators are included (green bar), the overall reach of global fit is not significantly degraded compared with the tree-level fit, and the reach on some of the operator coefficients are even improved.  This is because these operators already contribute at the tree level, so the number of operators is not increased, while their dependence on the observables are changed with the inclusion of loop effects.  

It should be noted that, as pointed out in \autoref{sec:Theory}, effects of the two bottom-quark dipole operators, $Q_{bW}$ and $Q_{bB}$ are non-negligible and should be included in the fit.  Even at the tree level, this introduces large flat directions with the operators $Q^{(+)}_{\varphi Q}$ and $Q_{\varphi b}$, as the four operators are mainly constrained by only two observables, $R_b$ and $A_b \,(A^b_{\rm FB})$.  Among them, we observe particular strong correlations among $Q_{bW}$, $Q_{bB}$ and $Q^{(+)}_{\varphi Q}$ ($>0.98$), while their correlations with $Q_{\varphi b}$ is relatively smaller ($<0.9$).  This is because the former three operators mainly contribute to $R_b$, while $Q_{\varphi b}$ modifies the $Zb_R\bar{b}_R$ coupling and is more sensitive to the asymmetry (see {\it e.g.} Ref.~\cite{Gori:2015nqa}).

%

Additional results are also provided in \autoref{sec:Appendix}.  \autoref{numerical_fit_result} shows both the marginalized bound with central values and individual bound. Figure~\ref{allfisher} shows the impact of different sets of measurements to the Wilson coefficients, which is given by the Fisher information.

\subsection{The $S\,T \, t\,b$ scenario}

\begin{figure}[t]
    \centering
            \includegraphics[width=0.4\linewidth]{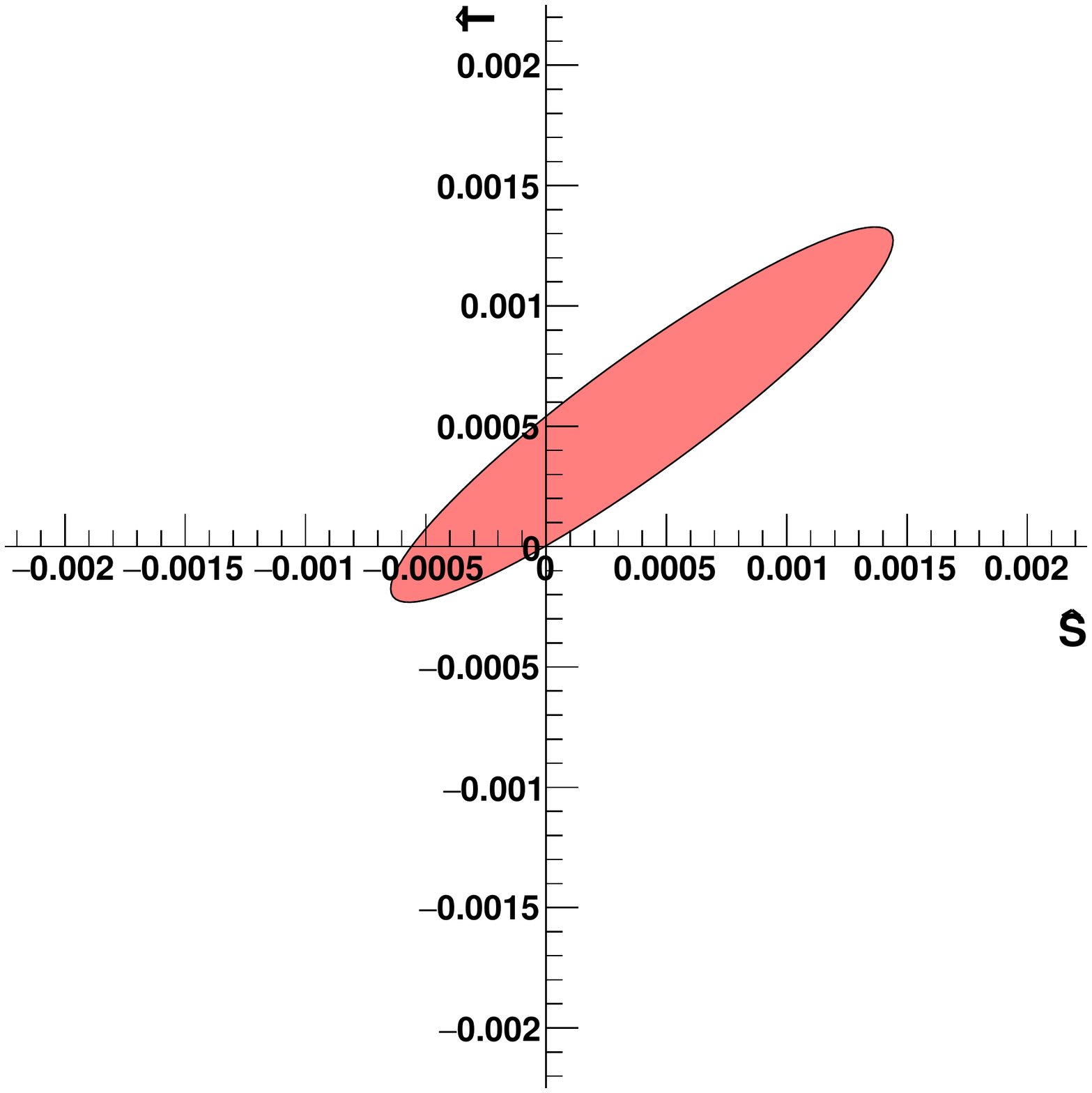}
        \label{STcontour}
    
    \caption{68\% CL preferred region of $\hat{S}$ and $\hat{T}$}
    \label{fig:ST}
\end{figure}

In additional to the general global-fitting framework, we also consider a special case denoted here as the ``$S\,T \, t\,b$'' scenario, where we assume that all new physics effects apart from the ones in \autoref{eq:Q1} can be parameterized by the two oblique parameters, $S$ and $T$~\cite{Peskin:1991sw}.  This scenario is motivatied by a large class of models with top/bottom partners, which generally mixes with 3rd generation quarks and could also contribute to oblique parameters at the one-loop order. 
For convenience, We work with the modified parameters $\hat{S}$ and $\hat{T}$~\cite{Barbieri:2004qk} which are related to the original ones by
\begin{equation}
\hat{S} = \frac{\alpha}{4 s^2_w} S \,, \hspace{1.5cm}
\hat{T} = \alpha T \,.
\end{equation}
$\hat{S}$, $\hat{T}$ are defined from \cite{Wells:2015uba}. The measurements considered are still the same as in \autoref{sec:ga}. 
For comparison, we first consider the two-parameter fit of $\hat{S}$ and $\hat{T}$. 
Their 68\% CL preferred region and correlation are shown in \autoref{fig:ST}. The numerical results of individual and  marginalized bound are:
%
\begin{equation}
\begin{matrix}
    \hat{S}=&(-2.8\pm2.5)\times10^{-4} \\
    \hat{T}=&(2.7\pm1.9)\times10^{-4}
\end{matrix}
\hspace{1cm} \mbox{(individual bound)}\,,
\end{equation}
and
%
\begin{equation}
\begin{matrix}
    \hat{S}=&(4.0\pm6.9)\times10^{-4} \\
    \hat{T}=&(5.5\pm5.1)\times10^{-4}
\end{matrix}
\hspace{1cm} \mbox{(marginalized bound)}\,.
\end{equation}
with the correlation:
\begin{equation}
    corr(\hat{S},\hat{T})=0.93
\end{equation}
These results are comparable with the ones in Ref.~\cite{Falkowski_2020} which basically uses the same set of measurements as we do. 

\begin{figure}[t]
    \centering
    \subfigure[]{
        \begin{minipage}[t]{0.5\linewidth}
        \centering
        \includegraphics[width=\linewidth]{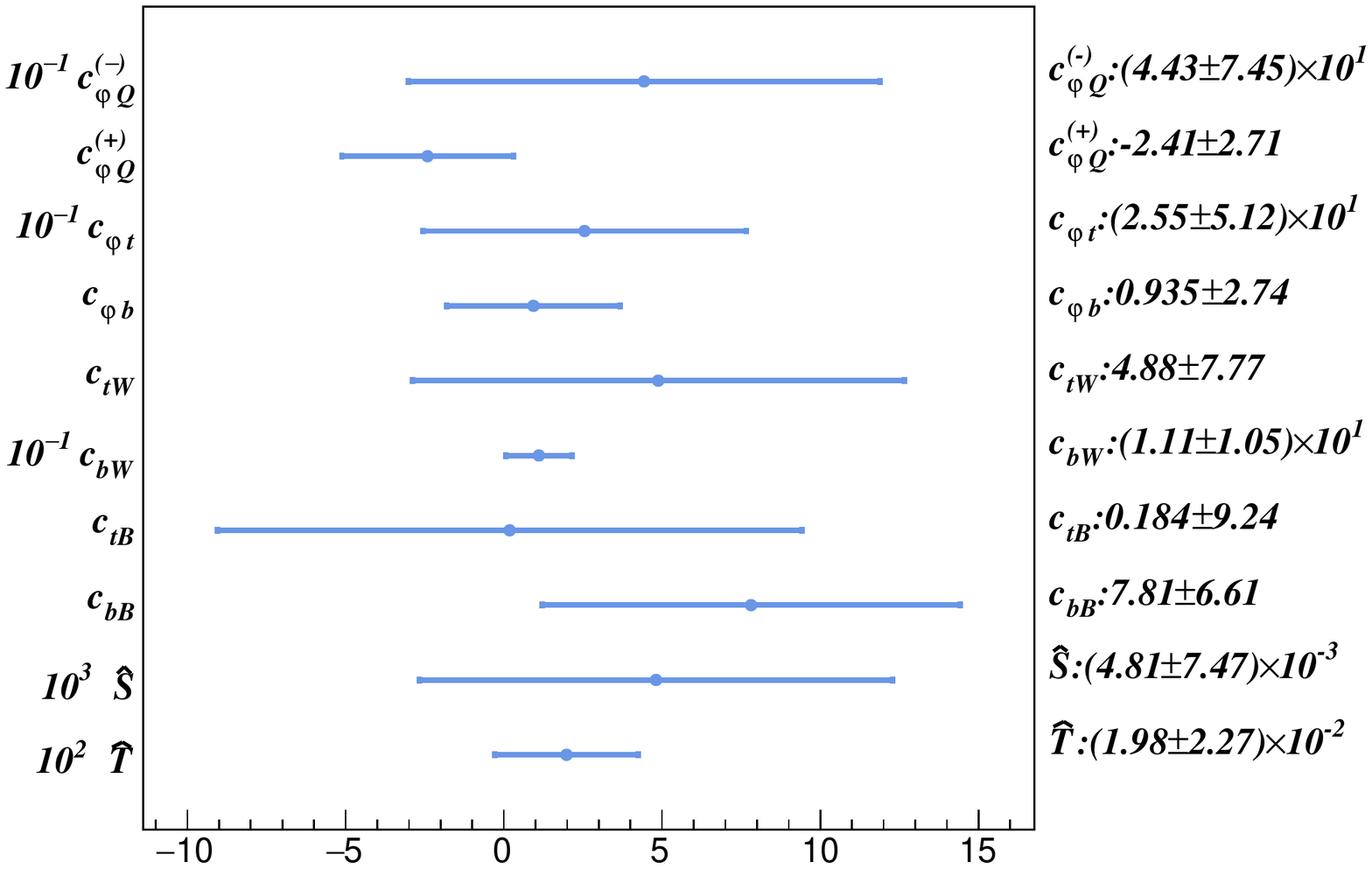}
        \end{minipage}
        \label{stprime_mar}
    }
    \subfigure[]{
        \begin{minipage}[t]{0.45\linewidth}
        \centering
        \includegraphics[width=\linewidth]{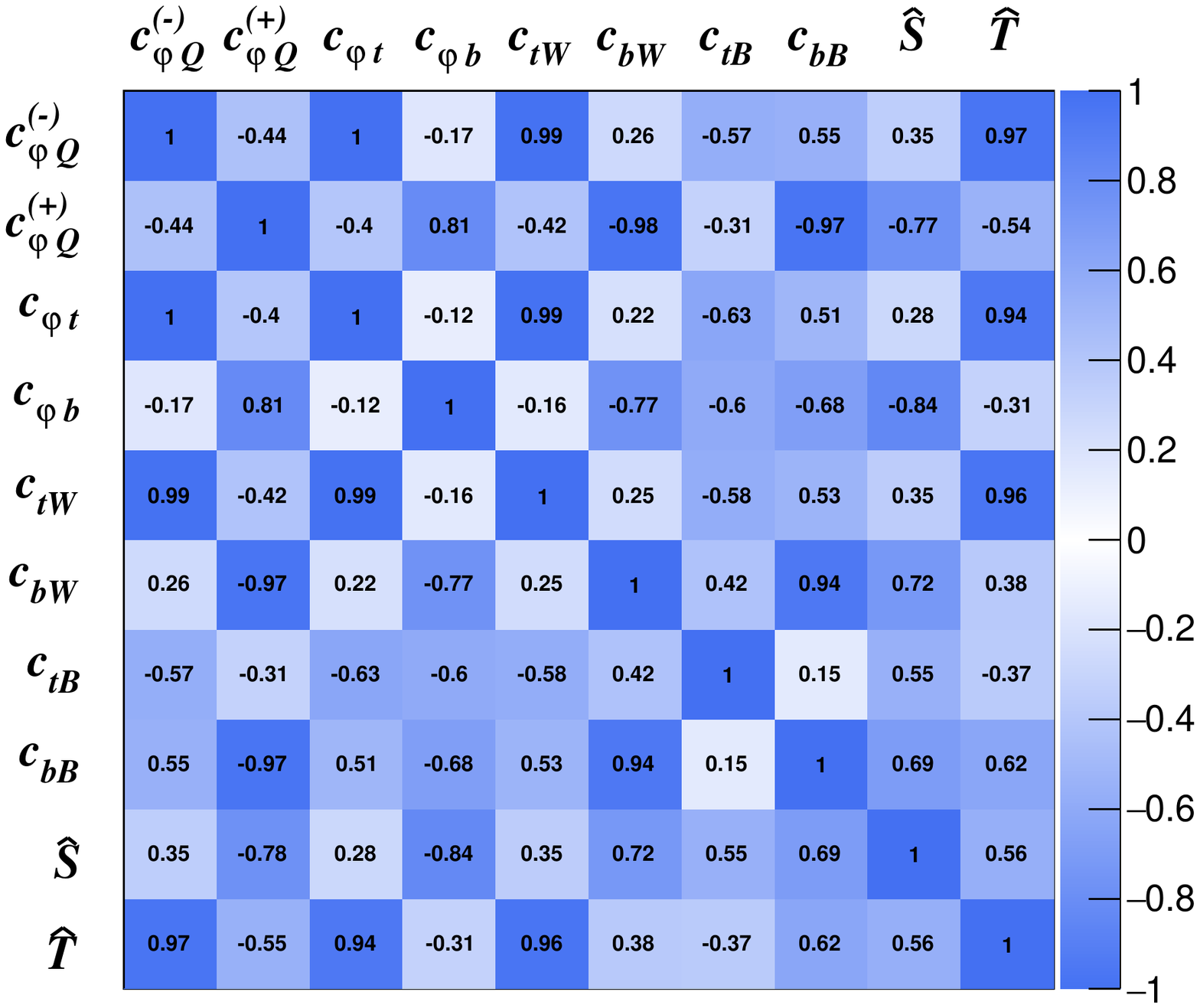}
        \end{minipage}
        \label{stloop_corr}
    }
    \caption{Fit results (Left: 68\% CL Marginalized bound, Right: Correlation matrix) of the $\hat{S}$, $\hat{T}$ parameters and the 8 operator coefficients in \autoref{eq:Q1}.}
    \label{stprime_results}
\end{figure}

With the inclusion of the 3rd-generation quark operators, we perform a 10-parameter fit, 
and the results are presented in \autoref{stprime_results}.  It should be noted that, part of the loop contributions of 3rd-generation quark operators are universal and can be absorbed in the original definition of the $S$ and $T$ parameters.  Here by $\hat{S}, \hat{T}$ we simply denote the contributions of the corresponding tree-level operators.  From \autoref{stprime_results}, we observe a significantly better reach on the top quark operators compared with the general case in \autoref{sec:ga}.   
On the other hand,  the constraining power to the $\hat{S}$ and $\hat{T}$ parameters with the electroweak data is significantly reduced from the 2-parameter case, as expected.  

One could also consider a more general set of universal correction~\cite{Wells:2015uba} (in addtion to the 3rd generation quark operators), which also requires additional measurements.  For instance, the $W$ and $Y$ parameters are strongly constrained by the high energy Drell-Yan measurement at the LHC~\cite{Farina:2016rws}. A detailed analysis in this direction is left for future studies.  

\subsection{Impacts of the new $W$-mass measurement at CDF}

\begin{figure}[htbp]
    \centering
   
        \centering
        \includegraphics[width=0.9\linewidth]{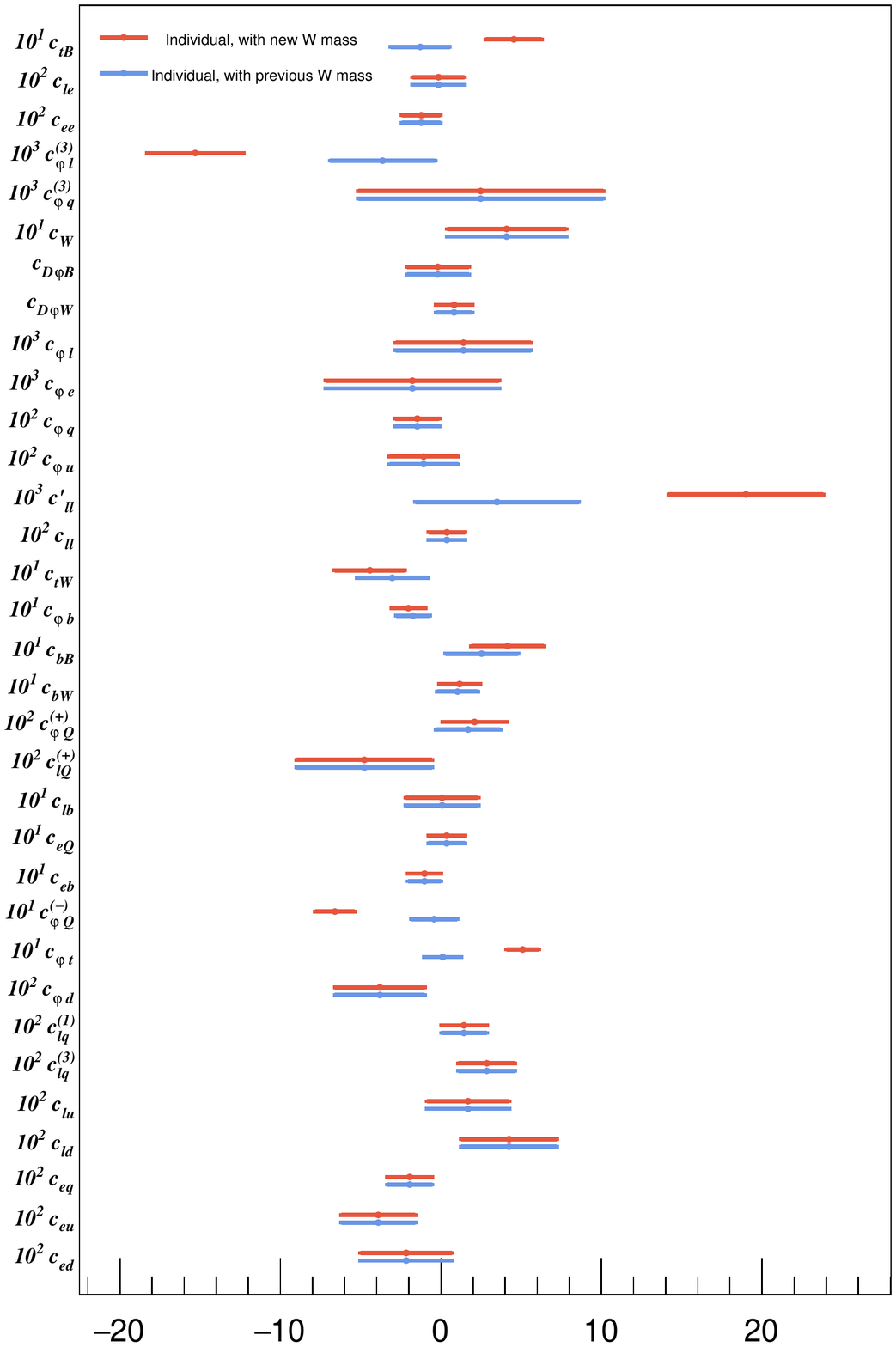}
        \label{34loop_newbasis_ind}
    
    \caption{Individual bound of coefficients of operators in modified Warsaw basis with previous (Blue) and new (Red) $W$ mass measurement. Here both the tree-level contribution and the loop contributions of all third-generation-quark operators are considered.}
    \label{newwmass_results}
\end{figure}

The CDF experiment recently announced their new measurement on the $W$ mass \cite{CDF:2022hxs}, which deviated away from SM prediction by seven standard deviations. This measurement will greatly affect the fit results. So in this part, the fit results with this new measured $W$ mass is shown in contrast to the ones with the previous measured value. \autoref{newwmass_results} shows the individual bound changes with the new measurement.

With the new CDF $W$-mass measurement, the results of the $\hat{S}$-$\hat{T}$ two-parameter fit are:
\begin{equation}
\begin{matrix}
    \hat{S}=&(-13.0\pm2.3)\times10^{-4} \\
    \hat{T}=&(10.9\pm1.6)\times10^{-4}
\end{matrix}
\hspace{1cm} \mbox{(individual bound)}\,,
\end{equation}
and
\begin{equation}
\begin{matrix}
    \hat{S}=&(13.2\pm6.8)\times10^{-4} \\
    \hat{T}=&(19.7\pm4.8)\times10^{-4}
\end{matrix}
\hspace{1cm} \mbox{(marginalized bound)}\,.
\end{equation}
%
%

\begin{figure}[htbp]
    \centering
   
        \centering
        \includegraphics[width=0.55\linewidth]{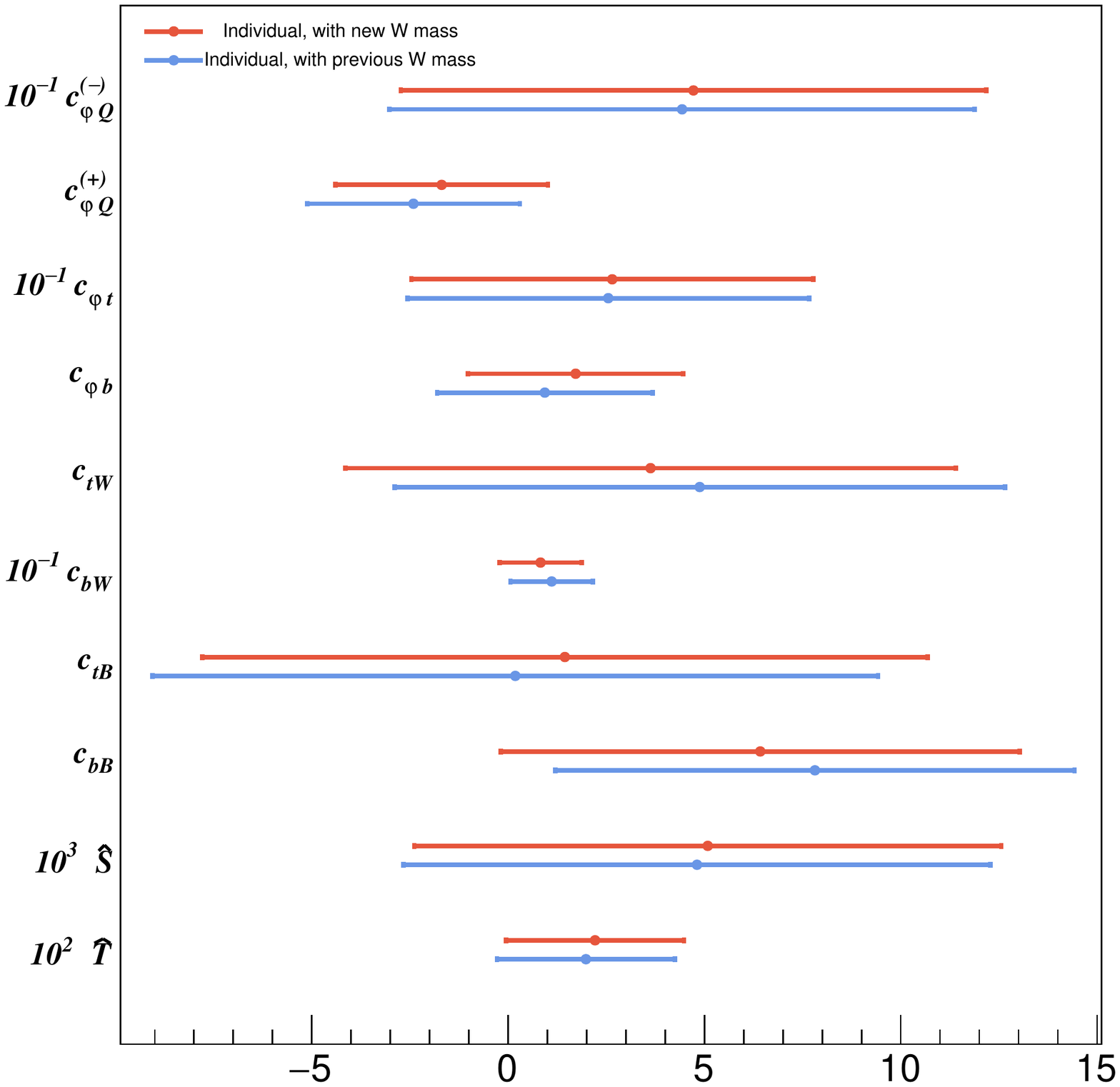}
    
    \caption{68\% CL Marginalized bound of coefficients of operators in ``$S\,T \, t\,b$'' scenario with the new $W$ mass measurement (red) and previous one (blue).}
    \label{stloop_newwmass_results}
\end{figure}
For the ``$S\,T \, t\,b$'' scenario, \autoref{stloop_newwmass_results} shows the marginalized bound with the new $W$ mass measurement. The impact of the shifted $W$ mass value is greatly absorbed by the 3rd-generation-quark operators.

There are several papers \cite{Bagnaschi:2022whn,Gu:2022htv,Fan:2022yly,Lu:2022bgw,Strumia:2022qkt,Gupta:2022lrt,Almeida:2022lcs,Balkin:2022glu,Endo:2022kiw,Asadi:2022xiy} have appeared that also study the impact of the new $m_W$ measurement on the new physics scenarios, especially on the SMEFT fit and oblique parameters. These results have a generally good agreement with our results. 
\section{Future Collider}
\label{sec:FutureCollider}
Recently, there have been several proposals of constructing future colliders which can reach higher luminosity and energies and will thus significantly increase the precision of the electroweak measurements. It is interesting to see how the improvement in measurements will affect the constraints of these EFT operators. As a result, the projections of electroweak observables from FCC-ee and CEPC are collected to conduct the chi-square fit. For the other two future collider proposals: ILC and CLIC, as Giga-Z proposals for them are still under discussion and currently there are little information about the expected performance of them on electroweak observables, they are not included in this study.

\begin{figure}[htbp]
    \centering
    \includegraphics[width=0.95\linewidth]{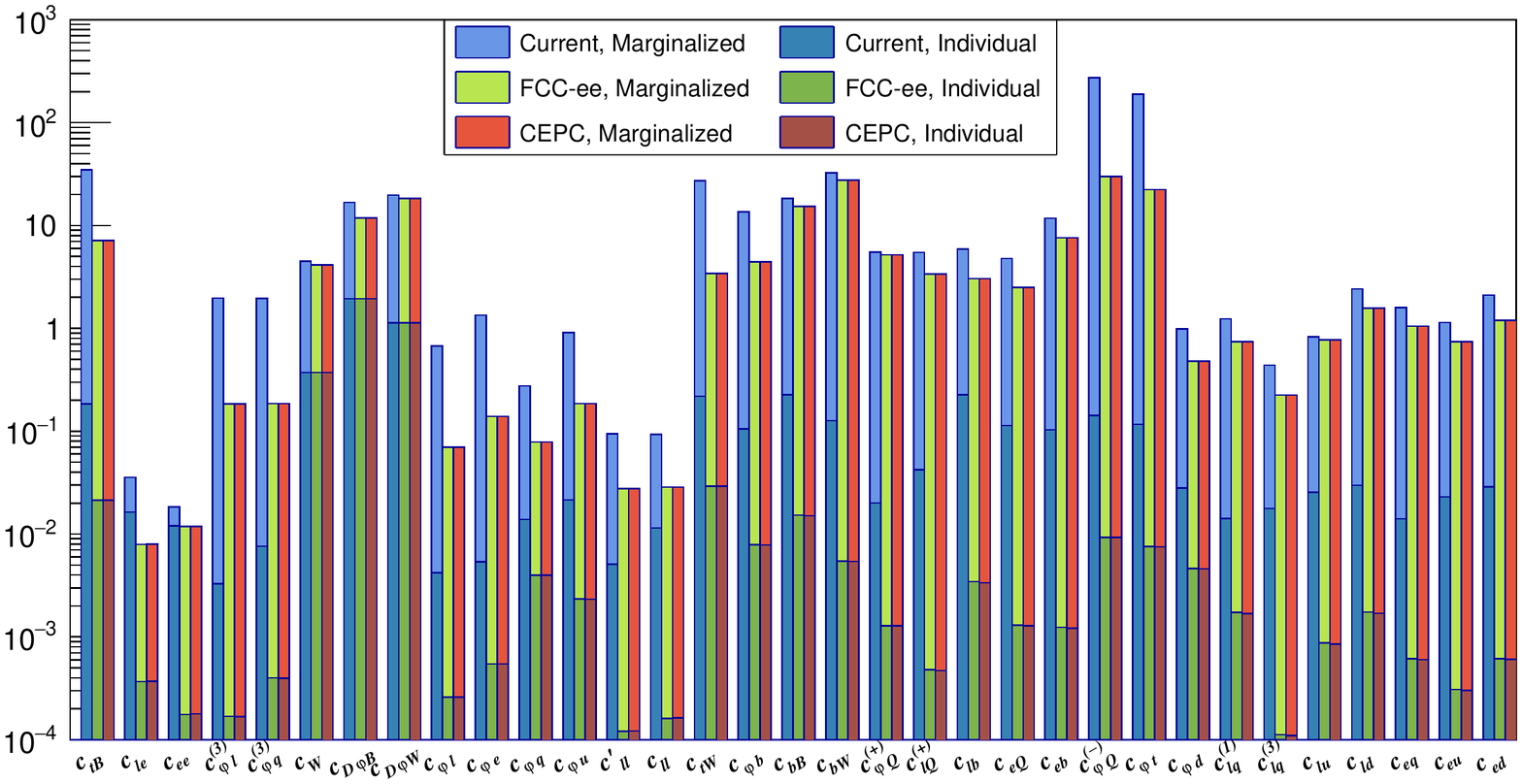}
    \label{future_newbasis}
    \caption{Comparison of the current precision of EW observables and the projections for CEPC and FCC-ee.}
\end{figure}

In table\,\ref{future_projections}, we list the comparison of current precision of electroweak observables and the projections for CEPC and FCC. As both two experiments are designed to be a tera-Z factory, the projections of observables at Z-pole are taken as a common value. For CEPC, the projections of luminosity at WW threshold and 240G$eV$ are taken from Ref.\cite{CEPCStudyGroup:2018ghi}. For FCC-ee, the predictions are taken from Ref.\cite{FCC:2018byv}. The projections of the constraints on the EFT operators involved in this study are shown in the figure\,\ref{future_newbasis}. In general, the future collider will greatly improve the constraining power on these EFT operators, most of the coefficients will get at least 10 times better constraints, both on individual and marginalized bound. Though the 3rd-generation quark operators can only be probed at loop level in current projections, they still got a much better constraints. For few operators, like $Q_W$ and $Q_{eq}$, as their constraints are mainly from $e^+e^-\to W^+W^-$ process or low energy coupling measurements, which lack the projection of their precision on future collider, their constraints have a relatively small improvement. However, it is believed that they can also benefit from the high precision measurements on future collider.  Finally, a high-energy lepton collider running at and above the top threshold will be able to measure the $e^+e^-\to t\bar{t}$ process to an unprecedented precision, providing the ultimate probe of the 3rd-generation quark operators~\cite{Durieux:2018tev, Durieux:2019rbz, Jung:2020uzh}.  The interplay between the top, Higgs and EW measurements in probing the 3rd-generation quark operators is an important topic for the future collider studies, which is beyond the scope of our current work.  

\begin{table}[htbp]
\centering
\begin{tabular}{|c|c|c|c|c|c|c|}
\hline
Observables & SM & Current Precision & CEPC & FCC-ee \\ \hline
$\Gamma_Z$ & 2.4942G$eV$ & 2.3M$eV$ & 0.1M$eV$ & 0.1M$eV$ \\ \hline 
$\sigma_{had}$ & 41.481nb & 0.037nb & 0.004nb & 0.004nb \\ \hline 
$R_e$ & 20.737 & 0.05 & 0.001 & 0.001 \\ \hline 
$R_\mu$ & 20.737 & 0.033 & 0.001 & 0.001 \\ \hline 
$R_\tau$ & 20.782 & 0.045 & 0.001 & 0.001 \\ \hline 
$R_b$ & 0.21582 & 0.00066 & 0.00006 & 0.00006 \\ \hline 
$R_c$ & 0.17221 & 0.003 & 0.00026 & 0.00026 \\ \hline 
$A_{FB}^b$ & 0.103 & 0.0003 & 0.0003 & 0.0003 \\ \hline 
$A_\tau$ & 0.1472 & 0.00216 & 0.0002 & 0.0002 \\ \hline 
$M_W$ & 80.379G$eV$ & 12M$eV$ & 1M$eV$ & 0.5M$eV$ \\ \hline 
$\Gamma_W$ & 2.085G$eV$ & 42M$eV$ & 2.8M$eV$ & 1.2M$eV$ \\ \hline 
$\sigma_{ee\to ff}^{160GeV}$& - & - & $\sqrt{\frac{\sigma_{SM}}{2.6ab^{-1}}}$ & $\sqrt{\frac{\sigma_{SM}}{12ab^{-1}}}$ \\ \hline
$\sigma_{ee\to ff}^{240GeV}$& - & - & $\sqrt{\frac{\sigma_{SM}}{5.6ab^{-1}}}$ & $\sqrt{\frac{\sigma_{SM}}{5ab^{-1}}}$ \\ \hline
$\sigma_{ee\to \mu\mu/\tau\tau, differential}^{240GeV}$& - & - & $\sqrt{\frac{\sigma_{SM}}{5.6ab^{-1}}}$ & $\sqrt{\frac{\sigma_{SM}}{5ab^{-1}}}$ \\ \hline
\end{tabular}
\caption{Comparison of the projections of electroweak observables for CEPC\cite{CEPCStudyGroup:2018ghi} and FCC-ee\cite{FCC:2018byv}.}
\label{future_projections}
\end{table}

\section{Conclusion}
\label{sec:Conclusion}

Precision measurements of the electroweak processes offer important probes to the physics beyond the Standard Model.  %
Many SMEFT analyses of the electroweak measurements focus on the tree-level contributions of dimension-6 operators.  %
However, given the outstanding precision of these measurements (especially at future lepton colliders), they could be sensitive to many important loop contributions of the new physics, which are not captured by a simple tree-level treatment. 
In this paper, we attempt to extend the tree-level framework by including the one-loop contributions of operators involving the third generation quarks, and study their impacts in a global analysis of the electroweak measurements.  This is motivated by many new physics scenarios where the 3rd generation quarks play a special role.   
We include the measurements of the $e^+e^-\to f\bar{f}$ processes at around the $Z$-pole and at several other energies, the measurements of $e^+e^-\to WW$ at LEP2, and a collection of low energy scattering processes.    
We find that the 3rd-generation quark operators, and especially the top quark ones, do have significant contributions to the electroweak processes.  In individual fits where only one operator coefficient is considered at a time, we obtain competitive reaches on these operators, which are all constrained to be at least around 1\,TeV, with the order-one coupling assumption.  On the other hand, in a global framework where all tree-level operator contributions are also included, the reaches on the operator coefficients become significantly worse, as it is difficult to separate the loop effects of the 3rd-generation quark operators from other tree-level effects.    
However, 
one should not conclude that the considerations of these loop contributions are meaningless in the SMEFT framework.  One important goal of the SMEFT is to provide a bridge between the experimental constraints and the  parameters in the UV model, and the likelihood from the SMEFT global analysis could be directly translated to the bounds on the UV model, even if the fit in SMEFT does not close.  
In a particular UV model, we usually expect a much smaller parameter space, and a global fit with loop effects become much more feasible.  As a demonstration, we performed a fit in a more constrained scenario where the only tree level contributions are parameterized by the two oblique parameters $S$ and $T$. In this case, much better constraints are obtained.  We also apply our analysis to the electroweak measurements at future lepton colliders.  
%
Another important finding of our results is that the tree-level contributions of the bottom dipole operators to the electroweak processes are non-negligible, and their effects are generally difficult to be separated from the modifications of the $Zb\bar{b}$ couplings. 

Our study is one of the many first steps towards a more complete loop-level SMEFT global analysis, for which many improvements are still needed.  Throughout our study, the loop effects of the 4-fermion operators involving the 3rd generation quarks are not considered.  These contributions could be comparable to the ones considered in our study and should in principle also be included.  However, they would introduce more degrees of freedom, and additional measurement are needed to discriminate their effects.  Similarly, it is also desirable to remove the flavor assumptions imposed in our study, which also significantly increases the size of the parameter space, and requires additional measurements~\cite{Breso-Pla:2021qoe}.  It is also important to study the complementarity between direct probes of the top quark operators, either at hadron colliders or a future lepton collider with higher center-of-mass energies, and the indirect ones studied here.  The 3rd generation quark operators also enters the Higgs processes, and a combined Higgs and electroweak analysis is particularly relevant for future lepton colliders in this framework.  An optimal-observable analysis of the $e^+e^- \to WW$ with also information in the $W$ decay angles was shown to be very useful in probing the corresponding tree-level operators~\cite{Diehl:1993br, DeBlas:2019qco}, and could be extended to also include loop effects.  However, this requires additional effort in calculating the one-loop contributions to the full differential cross section.  We leave these many possible extensions of our current analysis to future studies.

\acknowledgments

Yiming Liu, Cen Zhang and Jiayin Gu are supported by National Natural Science Foundation of China (NSFC) under grant No.~12035008. Yuhao Wang and Lei Zhang are supported by NSFC via grant No.~12122507. Cen Zhang, who was our collaborator, friend, and Yiming Liu's advisor, unexpectedly passed away while he was leading the work of this paper.  Jiayin Gu joined at a later stage to help complete this paper.



\appendix
\section{Appendix}
\label{sec:Appendix}
\subsection{List of the operators' contribution to operators}
\label{Appendix_A}
\begin{table}[htbp]
    \centering
    \resizebox{\linewidth}{!}{
    \begin{tabular}{|l|l|l|}
        \hline
        Processes&Observables&Wilson Coefficient\\ \hline
        Neutrino DIS and APV &$ g _ { L V } ^ { \nu e },g _ { L A } ^ { \nu e
        }$&$c^{(1)}_{\varphi l},c_{\varphi e},c_{ll} ,c_{le}$ \\   \cline{2-2}\cline{3-3}
        &\tabincell{l}{$g _ { A V } ^ { e u } + 2g _ { A V } ^ { e d } $\\
            $ 2g _ { A V } ^ { e u } - g _ { A V } ^ { e d }$ \\ $ 2g _ { V A } ^ { e u } -
            g _ { V A } ^ { e d }$}&\tabincell{l}{$c^{(1)}_{\varphi q}$,$c^{(3)}_{\varphi
                q},c_{\varphi u}$,$c_{\varphi d}$,$c^{(1)}_{\varphi l}$\\
            $c_{\varphi e}$,$c^{(1)}_{lq}$,$c^{(3)}_{lq}$,$c_{lu},c_{ld}$,
            $c_{qe}$,$c_{eu}$,$c_{ed}$}  \\
        \cline{2-2}\cline{3-3}
        &$g _ { V A } ^ { e e }$& \tabincell{l}{$c _ { \varphi l } ^ { ( 1 )
            },c_{\varphi e}, c _ { e e },c _ { l l}$}  \\ \cline{2-2}\cline{3-3}
        &2$\frac{g_{LV}^{\nu_{\mu}\mu,\mathrm{SM}}\delta
            g_{LV}^{\nu_{\mu}\mu}+g_{LA}^{\nu_{\mu}\mu,\mathrm{SM}}\delta
            g_{LA}^{\nu_{\mu}\mu}}{\left(g_{LV}^{\nu_{\mu}\mu,\mathrm{SM}}
            \right)^{2}+\left(g_{LA}^{\nu_{\mu}\mu,\mathrm{SM}} \right)^{2}}$
        &\tabincell{l}{$c_{\varphi l}^{(1)}$,$c_{\varphi e}$,$c_{le}$, $c_{ll}$}\\
        \cline{2-2}\cline{3-3}
        \hline
        $Z$-pole&\ $\Gamma_{Z}$& \tabincell{l}{$ c_ { \varphi l } ^ { ( 1 ) },c _ {
                \varphi e },c_ { \varphi q } ^ { ( 3) },c_ { \varphi q } ^ { ( 1 ) },c_{\varphi
                u},c _ { \varphi d },c_ { \varphi Q} ^ { ( + ) },c_ { \varphi b }, c_{bB}, c_{bW}$}  \\
        \cline{2-2}\cline{3-3}
        &\ $\sigma_{had}$& \tabincell{l}{$ c_ { \varphi l } ^ { ( 1 ) },c _ { \varphi e
            },c_ { \varphi q } ^ { ( 3) },c_ { \varphi q } ^ { ( 1 ) },c_{\varphi u},c _ {
                \varphi d },c_ { \varphi Q} ^ { ( + ) },c_ { \varphi b }, c_{bB}, c_{bW}$}  \\
        \cline{2-2}\cline{3-3}
        &\ $R_{e}$& \tabincell{l}{$ c_ { \varphi l } ^ { ( 1 ) },c _ { \varphi e },c_ {
                \varphi q } ^ { ( 3 ) },c_ { \varphi q } ^ { ( 1 ) },c_ { \varphi u },c_{\varphi
                d},c_ { \varphi Q} ^ { ( + ) },c _ {\varphi b }, c_{bB}, c_{bW}$}  \\ \cline{2-2}\cline{3-3}
        &\ $R_{\mu}$& \tabincell{l}{$ c_ { \varphi l } ^ { ( 1 ) },c _ { \varphi e },c_
            { \varphi q } ^ { ( 3 ) },c_ { \varphi q } ^ { ( 1 ) },c_ { \varphi u
            },c_{\varphi d},c_ { \varphi Q} ^ { ( + ) },c _ {\varphi b }, c_{bB}, c_{bW}$}  \\
        \cline{2-2}\cline{3-3}
        &\ $R_{\tau}$& \tabincell{l}{$ c_ { \varphi l } ^ { ( 1 ) },c _ { \varphi e },c_
            { \varphi q } ^ { ( 3 ) },c_ { \varphi q } ^ { ( 1 ) },c_ { \varphi u
            },c_{\varphi d},c_ { \varphi Q} ^ { ( + ) },c _ {\varphi b }, c_{bB}, c_{bW}$}  \\
        \cline{2-2}\cline{3-3}
        &\ $A^{o,e}_{FB}$& \tabincell{l}{$c _ { \varphi l } ^ { ( 1 ) },c_ { \varphi e
            }$}  \\ \cline{2-2}\cline{3-3}
        &\ $A^{o,\mu}_{FB}$& \tabincell{l}{$c _ { \varphi l } ^ { ( 1 ) },c_ { \varphi e
            }$}  \\ \cline{2-2}\cline{3-3}
        &\ $A^{o,\tau}_{FB}$&  \tabincell{l}{$c _ { \varphi l } ^ { ( 1 ) },c_ { \varphi
                e }$}  \\ \cline{2-2}\cline{3-3}
        &\ $R_{b}$& \tabincell{l}{$ c_ { \varphi q } ^ { ( 3 ) },c_ { \varphi q } ^ { (
                1 ) },c _ { \varphi u },c _ { \varphi d},c_ { \varphi Q} ^ { ( + ) },c_{\varphi
                b}, c_{bB}, c_{bW} $}  \\ \cline{2-2}\cline{3-3}
        &\ $R_{c}$&\tabincell{l}{$ c_ { \varphi q } ^ { ( 3 ) },c_ { \varphi q } ^ { ( 1
                ) },c _ { \varphi u },c _ { \varphi d},c_ { \varphi Q} ^ { ( + ) },c_{\varphi b}
            , c_{bB}, c_{bW}$}  \\ \cline{2-2}\cline{3-3}
        &$A^{b}_{FB}$& \tabincell{l}{$c_ { \varphi l } ^ { ( 1 ) },c_ { \varphi e },c_ {
                \varphi Q } ^ { ( + ) },c _ { \varphi b }, c_{bB}, c_{bW}$}  \\\cline{2-2}\cline{3-3}
        &$A^{c}_{FB}$& \tabincell{l}{$ c _ { \varphi l } ^ { ( 1 ) },c_ { \varphi e },c_
            { \varphi q } ^ { ( 1 ) },c _ { \varphi q } ^ { ( 3 ) },c _ { \varphi u },$}  \\
        \cline{2-2}\cline{3-3}
        &$A_{e}$& \tabincell{l}{$c _ { \varphi l } ^ { ( 1 ) },c_ { \varphi e }$}  \\
        \cline{2-2}\cline{3-3}
        &$A_{\mu}$& \tabincell{l}{$c _ { \varphi l } ^ { ( 1 ) },c_ { \varphi e }$}  \\
        \cline{2-2}\cline{3-3}
        &$A_{\tau}$& \tabincell{l}{$c _ { \varphi l } ^ { ( 1 ) },c_ { \varphi e }$}  \\
        \cline{2-2}\cline{3-3}
        &$A_{b}$& \tabincell{l}{$ c_ { \varphi Q } ^ { ( + ) },c _ { \varphi b }, c_{bB}, c_{bW}$}  \\
        \cline{2-2}\cline{3-3}
        &$A_{c}$& \tabincell{l}{$ c_ { \varphi q } ^ { ( 1 ) },c_ { \varphi q } ^ { ( 3
                ) },c _ { \varphi u }$}  \\ \cline{2-2}\cline{3-3}
        &$A_{s}$& \tabincell{l}{$ c_ { \varphi q } ^ { ( 1 ) },c_ { \varphi q } ^ { ( 3
                ) },c _ { \varphi d }$}  \\ \cline{2-2}\cline{3-3}
        \hline
        $W$-pole&$M_{W}$& \\
        \cline{2-2}\cline{3-3}
        &\tabincell{l}{$Br(W\to e\nu_{e})$$Br(W\to\mu\nu_{\mu})$\\$Br(W
            \to\tau\nu_{\tau})$$\Gamma_{W}$}& \tabincell{l}{$c _ { \varphi q } ^ { ( 3 ) }$}
        \\ \cline{2-2}\cline{3-3}
        \hline
        \tabincell{l}{$ee\to q\bar{q}$}&$\sigma_{eeqq}$& \tabincell{l}{$ c _ { \varphi l
            } ^ { ( 1 ) }$,$c _ { \varphi e }$,$ c _ { \varphi q } ^ { ( 3 ) }$,$c _ {
                \varphi q } ^ { ( 1 ) }$,$c _ { \varphi u }$,$c _ { \varphi d }$,$c _ { \varphi
                Q} ^ { ( + ) }$,$c_ { \varphi b }$,$c _ { l q } ^ { ( 1 ) }$,$c _ { l q } ^ { (
                3 ) }$,\\$c_{ l u }$,$c _ { l d }$,$c_ { l Q } ^ { ( + ) }$,$c _ { l b}$,$c _{ q
                e }$,$c_{eu}$,$c_ { e d }$,$c _ { eQ}$,$c _ { e b}, c_{bB}, c_{bW}$}  \\ \cline{2-2}\cline{3-3}
        \hline
        \tabincell{l}{$ee\to b\bar{b}$}& $\sigma_{eebb}$& \tabincell{l}{$ c _ { \varphi
                l } ^ { ( 1 ) },c _ { \varphi e },c _ { \varphi Q } ^ { ( + ) },c _ { \varphi b
            },c_ { l Q}^ { ( + ) },c _ { l b},c _ { eQ},c_ { e b}, c_{bB}, c_{bW}$}  \\
        \cline{2-2}\cline{3-3}
        \hline
        $ee\to c\bar{c}$&$\sigma_{eecc}$& \tabincell{l}{$c _ { \varphi l } ^ { ( 1 )
            }$,$c _ { \varphi e }$,$c _ { \varphi q } ^ { ( 1 ) }$,$c _ { \varphi q } ^ { (
                3 ) }$,$c_ { \varphi u }$,$c_{ l q } ^ { ( 1 ) }$,$c_ { l q } ^ { ( 3 ) }$,$c _
            { l u }$,$c_ { q e }$,$c_ { e u }$}  \\ \cline{2-2}\cline{3-3}
        \hline
        $ee\to \mu^{+}\mu^{-}$& $\sigma_{ee\mu\mu}$& \tabincell{l}{$c_ { \varphi l } ^ {
                ( 1 ) }$,$c_ { \varphi e }$,$c_ { ll}$,$c _ { l e }$,$c _ { e e } $}  \\
        \cline{2-2}\cline{3-3}
        \hline
        $ee\to\tau^{+}\tau^{-}$&$ \sigma_{ee\tau\tau}$&   \tabincell{l}{$c_ { \varphi l
            } ^ { ( 1 ) }$,$c_ { \varphi e }$,$c_ { ll}$,$c _ { l e }$,$c _ { e e } $}  \\
        \cline{2-2}\cline{3-3}
        \hline
        $ee\to e^{+}e^{-}$&$\sigma_{eeee}$&  \tabincell{l}{$c_ { \varphi l } ^ { ( 1 )
            }$,$c_ { \varphi e }$,$c_ { ll}$,$c _ { l e }$,$c _ { e e } $}  \\
        \cline{2-2}\cline{3-3}
        \hline
        $ee\to W^{+}W^{-}$&$\sigma_{eeww}$&\tabincell{l}{$c _ { \varphi l } ^ { ( 1 )
            }$,$c_ { \varphi e }$,$c _ { W }$}  \\ \cline{2-2}\cline{3-3}
        \hline 
    \end{tabular}}
    \caption{The list of corresponding coefficients of  type \uppercase\expandafter{\romannumeral1} operators that have contribution to different observables. Besides these coefficients, there are also four Wilson coefficients that will contribute to all observables: $c^{(3)}_{\varphi l}, c^{'}_{ll}, c_{\varphi D}, c_{\varphi WB}$.} 
    \label{tree_level_op}
\end{table}

\clearpage

\subsection{Additional fit results}
\label{add_fit_result}

\begin{table}[htbp]
    \centering
    \resizebox{\textwidth}{!}{
    \begin{tabular}{|c|c|c|c|c|}
    \hline
    Operators & Marginalized, tree-only & Marginalized, b loop considered & Marginalized, t/b loop considered & Individual, t/b loop considered \\ \hline
    $c_{tB}$ & - & - & $30 \pm 35$ & $-0.13 \pm 0.18$  \\ \hline
$c_{le}$ & $0.0019 \pm 0.018$ & $0.0014 \pm 0.018$ & $0.036 \pm 0.036$ & $-0.0015 \pm 0.016$  \\ \hline
$c_{ee}$ & $-0.019 \pm 0.014$ & $-0.019 \pm 0.014$ & $-0.0082 \pm 0.018$ & $-0.012 \pm 0.012$  \\ \hline
$c_{\varphi l}^{(3)}$ & $-0.023 \pm 0.023$ & $-0.027 \pm 0.018$ & $-0.42 \pm 2$ & $-0.0036 \pm 0.0033$  \\ \hline
$c_{\varphi q}^{(3)}$ & $-0.079 \pm 0.07$ & $-0.11 \pm 0.074$ & $-0.47 \pm 2$ & $0.0025 \pm 0.0076$  \\ \hline
$c_{W}$ & $-7.9 \pm 5.1$ & $-1.8 \pm 3.7$ & $-4.5 \pm 4.5$ & $0.41 \pm 0.37$  \\ \hline
$c_{D\varphi B}$ & $-13 \pm 10$ & $-7.9 \pm 11$ & $-12 \pm 17$ & $-0.2 \pm 1.9$  \\ \hline
$c_{D\varphi W}$ & $35 \pm 21$ & $12 \pm 17$ & $23 \pm 20$ & $0.82 \pm 1.1$  \\ \hline
$c_{\varphi l}$ & $0.024 \pm 0.013$ & $0.021 \pm 0.013$ & $-0.26 \pm 0.67$ & $0.0014 \pm 0.0042$  \\ \hline
$c_{\varphi e}$ & $0.0051 \pm 0.014$ & $-2.5e-05 \pm 0.013$ & $-0.57 \pm 1.3$ & $-0.0018 \pm 0.0054$  \\ \hline
$c_{\varphi q}$ & $-0.042 \pm 0.13$ & $-0.053 \pm 0.14$ & $0.085 \pm 0.28$ & $-0.015 \pm 0.014$  \\ \hline
$c_{\varphi u}$ & $0.086 \pm 0.17$ & $0.11 \pm 0.17$ & $0.49 \pm 0.91$ & $-0.011 \pm 0.021$  \\ \hline
$c'_{ll}$ & $-0.043 \pm 0.03$ & $-0.055 \pm 0.029$ & $-0.069 \pm 0.094$ & $0.0035 \pm 0.0051$  \\ \hline
$c_{ll}$ & $0.054 \pm 0.033$ & $0.066 \pm 0.031$ & $0.1 \pm 0.093$ & $0.0037 \pm 0.011$  \\ \hline
$c_{tW}$ & - & - & $4.2 \pm 27$ & $-0.3 \pm 0.22$  \\ \hline
$c_{\varphi b}$ & $-0.34 \pm 0.62$ & $-0.27 \pm 0.63$ & $-12 \pm 14$ & $-0.17 \pm 0.11$  \\ \hline
$c_{bB}$ & $10 \pm 15$ & $9.2 \pm 14$ & $12 \pm 18$ & $0.25 \pm 0.23$  \\ \hline
$c_{bW}$ & $16 \pm 27$ & $9.7 \pm 25$ & $17 \pm 32$ & $0.1 \pm 0.13$  \\ \hline
$c_{\varphi Q}^{(+)}$ & $-3.5 \pm 5.4$ & $-2.3 \pm 5$ & $-6.1 \pm 5.5$ & $0.017 \pm 0.02$  \\ \hline
$c_{lQ}^{(+)}$ & $-1.5 \pm 2.1$ & $-2 \pm 2.4$ & $-4.7 \pm 5.5$ & $-0.048 \pm 0.042$  \\ \hline
$c_{lb}$ & $1.2 \pm 2.3$ & $1.4 \pm 2.5$ & $-1.2 \pm 5.9$ & $0.0074 \pm 0.23$  \\ \hline
$c_{eQ}$ & $0.64 \pm 1.8$ & $0.56 \pm 1.8$ & $-1.5 \pm 4.8$ & $0.035 \pm 0.11$  \\ \hline
$c_{eb}$ & $3.6 \pm 5$ & $5.1 \pm 5.6$ & $11 \pm 12$ & $-0.1 \pm 0.1$  \\ \hline
$c_{\varphi Q}^{(-)}$ & - & - & $31 \pm 2.7e+02$ & $-0.043 \pm 0.14$  \\ \hline
$c_{\varphi t}$ & - & - & $3.8 \pm 1.9e+02$ & $0.011 \pm 0.12$  \\ \hline
$c_{\varphi d}$ & $-0.7 \pm 0.8$ & $-0.92 \pm 0.86$ & $-0.69 \pm 0.99$ & $-0.038 \pm 0.028$  \\ \hline
$c_{lq}^{(1)}$ & $2.3 \pm 1.2$ & $2.7 \pm 1.2$ & $2.6 \pm 1.2$ & $0.014 \pm 0.014$  \\ \hline
$c_{lq}^{(3)}$ & $0.74 \pm 0.43$ & $0.88 \pm 0.43$ & $0.84 \pm 0.44$ & $0.029 \pm 0.018$  \\ \hline
$c_{lu}$ & $-1.1 \pm 0.83$ & $-1.3 \pm 0.82$ & $-1.3 \pm 0.83$ & $0.017 \pm 0.026$  \\ \hline
$c_{ld}$ & $-4.8 \pm 2.4$ & $-5.8 \pm 2.4$ & $-5.2 \pm 2.4$ & $0.042 \pm 0.03$  \\ \hline
$c_{eq}$ & $2.6 \pm 1.6$ & $3 \pm 1.6$ & $2.8 \pm 1.6$ & $-0.019 \pm 0.014$  \\ \hline
$c_{eu}$ & $-2.2 \pm 1.1$ & $-2.6 \pm 1.1$ & $-2.4 \pm 1.1$ & $-0.039 \pm 0.023$  \\ \hline
$c_{ed}$ & $-3.6 \pm 2.1$ & $-4.2 \pm 2$ & $-4.1 \pm 2.1$ & $-0.022 \pm 0.029$  \\ \hline
    \end{tabular}}
    \caption{Numerical fit result of the coefficients of operators in modified Warsaw basis for three different scenarios. The error bound is given in 68\%CL.}
    \label{numerical_fit_result}
\end{table}

The Fisher information of set $i$ have on coefficient $c_j$ is calculated by:
\begin{equation}   
    f_i=\frac{\frac{\partial^2 \chi^2_i}{\partial c_j^2}}{\frac{\partial^2 \chi^2_{all}}{\partial c_j^2}},
\end{equation}
where $\chi^2_i$ indicated the chi-square calculated with only the data of set $i$ and $\chi^2_{all}$ indicates the chi-square with all sets of measurements.

\begin{figure}[htbp]
    \centering
    \includegraphics[width=0.7\linewidth]{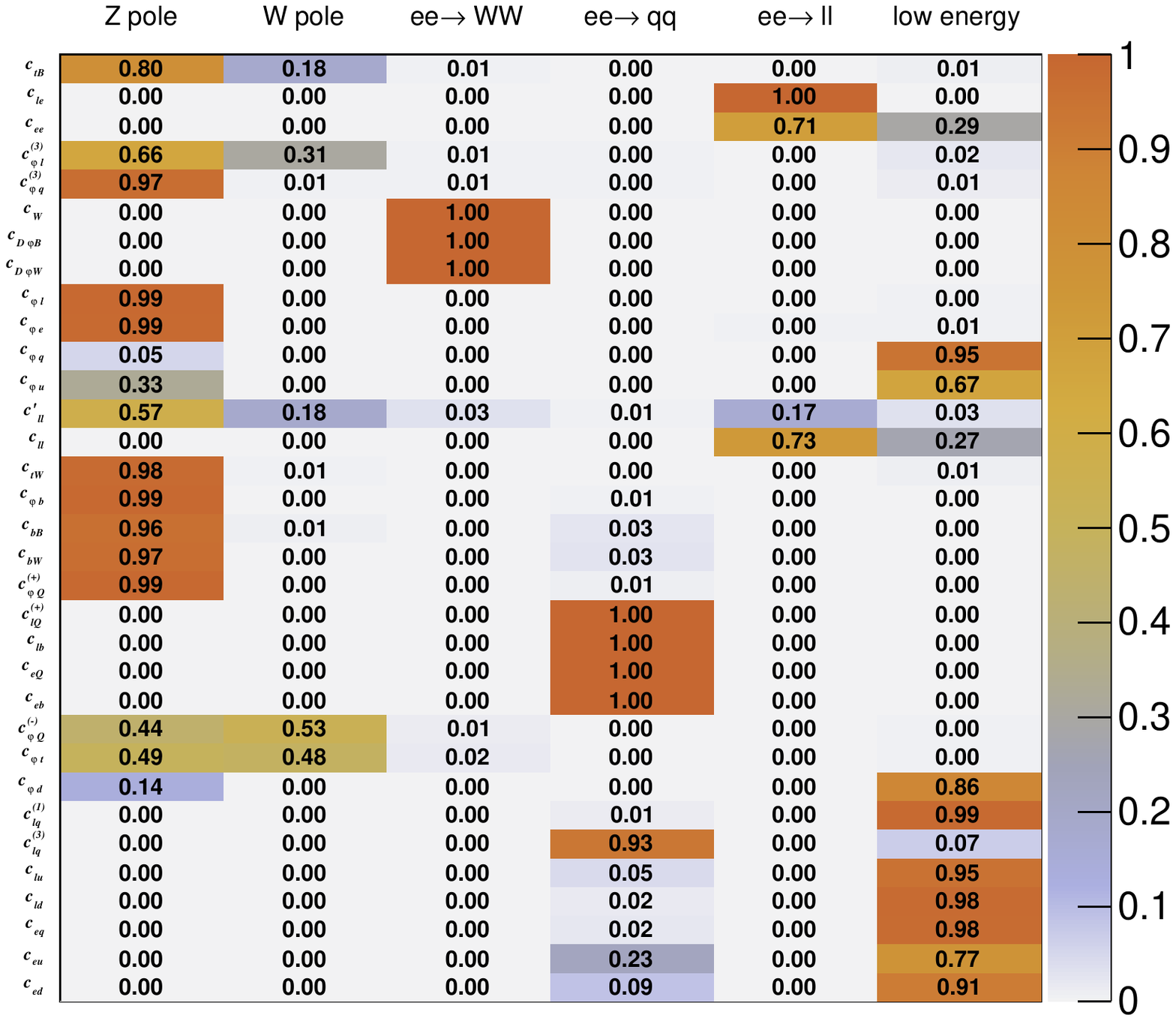}
    \caption{Impacts of different set of measurement to the Wilson coefficients in Warsaw basis, measured by the Fisher information. The larger the number is, the bigger impact of the set have on the constraints of Wilson coefficients. }
    \label{allfisher}
\end{figure}



\subsection{Definition of renormalized parameters}
\label{def_renormalized_par}

\begin{align}
m^{2}_{W^{*}}(q^{2})&=(1-Z_{W})q^{2}+Z_{W}\bigg(m^{2} _{WO}+\Pi_{WW}(q^{2})\bigg)\label{mw}\\
m^{2}_{Z^{*}}(q^{2})&=(1-Z_{Z})q^{2}+Z_{Z}\bigg(m^{2} _{ZO}+\Pi_{ZZ}(q^{2})\bigg)\\
Z_{W}&=1+\frac{\mathrm{d}\Pi_{WW}(q^{2})}{\mathrm{d}q^{2}}|_{q^{2}=m^{2}_{W}}\\
Z_{Z}&=1+\frac{\mathrm{d}\Pi_{ZZ}(q^{2})}{\mathrm{d}q^{2}}|_{q^{2}=m^{2}_{Z}}\\
Z_{W^{*}}(q^{2})&=1+\frac{\mathrm{d}\Pi_{WW}(q^{2})}{\mathrm{d}q^{2}}|_{q^{2}=m^{2}_{W}}-\Pi^{'}_{\gamma\gamma}(q^{2})-\frac{c_{W}}{s_{W}}\Pi^{'}_{\gamma Z}(q^{2})\\
Z_{Z^{*}}(q^{2})&=1+\frac{\mathrm{d}\Pi_{ZZ}(q^{2})}{\mathrm{d}q^{2}}|_{q^{2}=m^{2}_{Z}}-\Pi^{'}_{\gamma\gamma}(q^{2})-\frac{c^{2}_{W}-s^{2}_{W}}{s_{W}c_{W}}\Pi^{'}_{\gamma Z}(q^{2})\\
s^{2}_{W^{*}}(q^{2})&=s^{2} _{WO}-s_{W}c_{W}\Pi^{'}_{\gamma Z}(q^{2})\\
e^{2}_{*}(q^{2})&=e^{2} _{O}+e^{2}\Pi^{'}_{\gamma\gamma}(q^{2})\\
\Pi^{'}_{XY}(q^{2})&=\frac{\big(\Pi_{XY}(q^{2})-\Pi_{XY}(0)\big)}{q^{2}}\\
\Pi_{XY}(q^{2})&=\sum_{i} c_{i}\Pi^{(i)}_{XY}(q^{2})
\end{align}

\begin{itemize}         
    \item{$Q_{\varphi q}^{33(3)}$}
    \begin{eqnarray}
    \Pi^{(1)}_{WW}&=&-N_c\frac{g^2}{4\pi^2}\frac{v^2}{\Lambda^2}
    \left[
    \left(\frac{1}{6}q^2-\frac{1}{4}(m_t^2+m_b^2)\right)E
    \right.\nonumber\\&&\left.
    -q^2b_2(m_t^2,m_b^2,q^2)+\frac{1}{2}\left(m_b^2b_1(m_t^2,m_b^2,q^2)+m_t^2b_1(m_b^2,m_t^2,q^2)\right)
    \right]
    \\
    \Pi^{(1)}_{ZZ}&=&-N_c\frac{g^2}{\cos^2\theta_W}\frac{1}{4\pi^2}\frac{v^2}{\Lambda^2}
    \left[
    \left(\frac{1}{6}(1-\sin^2\theta_W)q^2-\frac{1}{4}(m_t^2+m_b^2)\right)E
    \right.\nonumber\\&&\left.
    -q^2\left(\left(\frac{1}{2}-\frac{2}{3}\sin^2\theta_W\right)b_2(m_t^2,m_t^2,q^2)+\left(\frac{1}{2}-\frac{1}{3}\sin^2\theta_W\right)b_2(m_b^2,m_b^2,q^2)\right)
    \right.\nonumber\\&&\left.
    +\frac{1}{4}\left(m_t^2b_0(m_t^2,m_t^2,q^2)+m_b^2b_0(m_b^2,m_b^2,q^2)\right)
    \right]
    \\
    \Pi^{(1)}_{\gamma\gamma}&=&0
    \\
    \Pi^{(1)}_{\gamma
        Z}&=&-N_cg^2\frac{\sin\theta_W}{\cos\theta_W}\frac{1}{8\pi^2}\frac{v^2}{\Lambda^2}
    \left[\frac{1}{6}E-\frac{2}{3}b_2(m_t^2,m_t^2,q^2)-\frac{1}{3}b_2(m_b^2,m_b^2,q^2)\right]q^2
    \end{eqnarray}
    \item{$Q_{\varphi q}^{33(1)}$}
    \begin{eqnarray}
    \Pi^{(2)}_{WW}&=&0
    \\
    \Pi^{(2)}_{ZZ}&=&N_c\frac{g^2}{\cos^2\theta_W}\frac{1}{4\pi^2}\frac{v^2}{\Lambda^2}
    \left[-\left(\frac{1}{4}m_t^2-\frac{1}{4}m_b^2+\frac{1}{18}q^2\sin^2\theta_W\right)E
    \right.\nonumber\\&&\left.
    -q^2\left(\left(\frac{1}{2}-\frac{2}{3}\sin^2\theta_W\right)b_2(m_t^2,m_t^2,q^2)
    -\left(\frac{1}{2}-\frac{1}{3}\sin^2\theta_W\right)b_2(m_b^2,m_b^2,q^2)\right)
    \right.\nonumber\\&&\left.
    +\frac{1}{4}\left(m_t^2b_0(m_t^2,m_t^2,q^2)-m_b^2b_0(m_b^2,m_b^2,q^2)\right)
    \right]
    \\
    \Pi^{(2)}_{\gamma\gamma}&=&0
    \\
    \Pi^{(2)}_{\gamma
        Z}&=&N_cg^2\frac{\sin\theta_W}{\cos\theta_W}\frac{1}{8\pi^2}\frac{v^2}{\Lambda^2}
    \left[\frac{1}{18}E-\frac{2}{3}b_2(m_t^2,m_t^2,q^2)+\frac{1}{3}b_2(m_b^2,m_b^2,q^2)\right]q^2
    \end{eqnarray}
    \item{$Q^{33}_{\varphi u}$}
    \begin{eqnarray}
    \Pi^{(3)}_{WW}&=&0
    \\
    \Pi^{(3)}_{ZZ}&=&N_c\frac{g^2}{\cos^2\theta_W}\frac{1}{4\pi^2}\frac{v^2}{\Lambda^2}
    \left[\left(\frac{1}{4}m_t^2-\frac{1}{9}q^2\sin^2\theta_W\right)E
    \right.\nonumber\\&&\left.
    -\left(\frac{1}{4}m_t^2b_0(m_t^2,m_t^2,q^2)-\frac{2}{3}q^2\sin^2\theta_Wb_2(m_t^2,m_t^2,q^2)\right)
    \right]
    \\
    \Pi^{(3)}_{\gamma\gamma}&=&0
    \\
    \Pi^{(3)}_{\gamma
        Z}&=&N_cg^2\frac{\sin\theta_W}{\cos\theta_W}\frac{1}{12\pi^2}\frac{v^2}{\Lambda^2}
    \left(\frac{1}{6}E-b_2(m_t^2,m_t^2,q^2)\right)q^2
    \end{eqnarray}
    \item{$Q^{33}_{\varphi d}$}
    \begin{eqnarray}
    \Pi^{(4)}_{WW}&=&0
    \\
    \Pi^{(4)}_{ZZ}&=&N_c\frac{g^2}{\cos^2\theta_W}\frac{1}{4\pi^2}\frac{v^2}{\Lambda^2}
    \left[-\left(\frac{1}{4}m_b^2-\frac{1}{18}q^2\sin^2\theta_W\right)E
    \right.\nonumber\\&&\left.
    +\left(\frac{1}{4}m_b^2b_0(m_b^2,m_b^2,q^2)-\frac{1}{3}q^2\sin^2\theta_Wb_2(m_b^2,m_b^2,q^2)\right)
    \right]
    \\
    \Pi^{(4)}_{\gamma\gamma}&=&0
    \\
    \Pi^{(4)}_{\gamma
        Z}&=&-N_cg^2\frac{\sin\theta_W}{\cos\theta_W}\frac{1}{24\pi^2}\frac{v^2}{\Lambda^2}
    \left(\frac{1}{6}E-b_2(m_b^2,m_b^2,q^2)\right)q^2
    \end{eqnarray}
    \item{$Q^{33}_{\varphi ud}$}
    \begin{eqnarray}
    \Pi^{(5)}_{WW}&=&-N_cg^2\frac{1}{16\pi^2}\frac{v^2}{\Lambda^2}m_tm_b\left(E-b_0(m_t^2,m_b^2,q^2)\right)
    \label{phiphi}
    \\
    \Pi^{(5)}_{ZZ}&=&0
    \\
    \Pi^{(5)}_{\gamma\gamma}&=&0
    \\
    \Pi^{(5)}_{\gamma Z}&=&0
    \end{eqnarray}
    \item{$Q^{33}_{uW}$}
    \begin{eqnarray}
    \Pi^{(6)}_{WW}&=&-N_cg\frac{\sqrt{2}}{4\pi^2}\frac{vm_t}{\Lambda^2}\left(\frac{1}{2}E-b_1(m_b^2,m_t^2,q^2)\right)q^2
    \\
    \Pi^{(6)}_{ZZ}&=&-N_cg\frac{\sqrt{2}}{4\pi^2}\frac{vm_t}{\Lambda^2}\left(\frac{1}{2}-\frac{4}{3}\sin^2\theta_W\right)
    \left(E-b_0(m_t^2,m_t^2,q^2)\right)q^2
    \\
    \Pi^{(6)}_{\gamma\gamma}&=&-N_cg\frac{\sqrt{2}}{4\pi^2}\frac{vm_t}{\Lambda^2}\frac{4}{3}\sin^2\theta_W\left(E-b_0(m_t^2,m_t^2,q^2)\right)q^2
    \\
    \Pi^{(6)}_{\gamma
        Z}&=&-N_cg\frac{\sqrt{2}}{4\pi^2}\frac{vm_t}{\Lambda^2}\frac{\sin\theta_W}{\cos\theta_W}
    \left(\frac{11}{12}-\frac{4}{3}\sin^2\theta_W\right)\left(E-b_0(m_t^2,m_t^2,q^2)\right)q^2
    \end{eqnarray}
    \item{$Q^{33}_{dW}$}
    \begin{eqnarray}
    \Pi^{(7)}_{WW}&=&-N_cg\frac{\sqrt{2}}{4\pi^2}\frac{vm_b}{\Lambda^2}\left(\frac{1}{2}E-b_1(m_t^2,m_b^2,q^2)\right)q^2
    \\
    \Pi^{(7)}_{ZZ}&=&-N_cg\frac{\sqrt{2}}{4\pi^2}\frac{vm_b}{\Lambda^2}\left(\frac{1}{2}-\frac{2}{3}\sin^2\theta_W\right)
    \left(E-b_0(m_b^2,m_b^2,q^2)\right)q^2
    \\
    \Pi^{(7)}_{\gamma\gamma}&=&-N_cg\frac{\sqrt{2}}{4\pi^2}\frac{vm_b}{\Lambda^2}\frac{2}{3}\sin^2\theta_W\left(E-b_0(m_b^2,m_b^2,q^2)\right)q^2
    \\
    \Pi^{(7)}_{\gamma
        Z}&=&-N_cg\frac{\sqrt{2}}{4\pi^2}\frac{vm_b}{\Lambda^2}\frac{\sin\theta_W}{\cos\theta_W}
    \left(\frac{7}{12}-\frac{2}{3}\sin^2\theta_W\right)\left(E-b_0(m_b^2,m_b^2,q^2)\right)q^2
    \end{eqnarray}
    \item{$Q^{33}_{uB}$}
    \begin{eqnarray}
    \Pi^{(8)}_{WW}&=&0
    \\
    \Pi^{(8)}_{ZZ}&=&N_cg\frac{\sqrt{2}}{4\pi^2}\frac{vm_t}{\Lambda^2}\frac{\sin\theta_W}{\cos\theta_W}\left(\frac{1}{2}-\frac{4}{3}\sin^2\theta_W\right)
    \left(E-b_0(m_t^2,m_t^2,q^2)\right)q^2
    \\
    \Pi^{(8)}_{\gamma\gamma}&=&-N_cg\frac{\sqrt{2}}{4\pi^2}\frac{vm_t}{\Lambda^2}\frac{4}{3}\sin\theta_W
    \cos\theta_W\left(E-b_0(m_t^2,m_t^2,q^2)\right)q^2
    \\
    \Pi^{(8)}_{\gamma Z}&=&-N_cg\frac{\sqrt{2}}{4\pi^2}\frac{vm_t}{\Lambda^2}
    \left(\frac{1}{4}-\frac{4}{3}\sin^2\theta_W\right)\left(E-b_0(m_t^2,m_t^2,q^2)\right)q^2
    \end{eqnarray}
    \item{$Q^{33}_{dB}$}
    \begin{eqnarray}
    \Pi^{(9)}_{WW}&=&0
    \\
    \Pi^{(9)}_{ZZ}&=&-N_cg\frac{\sqrt{2}}{4\pi^2}\frac{vm_b}{\Lambda^2}\frac{\sin\theta_W}{\cos\theta_W}\left(\frac{1}{2}-\frac{2}{3}\sin^2\theta_W\right)
    \left(E-b_0(m_b^2,m_b^2,q^2)\right)q^2
    \\
    \Pi^{(9)}_{\gamma\gamma}&=&N_cg\frac{\sqrt{2}}{4\pi^2}\frac{vm_b}{\Lambda^2}\frac{2}{3}\sin\theta_W
    \cos\theta_W\left(E-b_0(m_b^2,m_b^2,q^2)\right)q^2
    \\
    \Pi^{(9)}_{\gamma Z}&=&N_cg\frac{\sqrt{2}}{4\pi^2}\frac{vm_b}{\Lambda^2}
    \left(\frac{1}{4}-\frac{2}{3}\sin^2\theta_W\right)\left(E-b_0(m_b^2,m_b^2,q^2)\right)q^2
    \end{eqnarray}
\end{itemize}
Where $\theta_W$ is the weak angle, $N_c=3$ is the number of colors.
$E=\frac{2}{4-d}-\gamma+\ln 4\pi$, and the functions $b_i$ are given by
\begin{eqnarray}
	&&b_0(m_1^2,m_2^2,q^2)=\int_0^1\ln\frac{(1-x)m_1^2+xm_2^2-x(1-x)q^2}{\mu^2}dx,\\
	&&b_1(m_1^2,m_2^2,q^2)=\int_0^1x\ln\frac{(1-x)m_1^2+xm_2^2-x(1-x)q^2}{\mu^2}dx,\\
	&&b_2(m_1^2,m_2^2,q^2)=\int_0^1x(1-x)\ln\frac{(1-x)m_1^2+xm_2^2-x(1-x)q^2}{\mu^2}dx,
\end{eqnarray}
where $\mu$ is the 't Hooft mass. They have the following analytical expressions:
\begin{eqnarray}
	b_0(m_1^2,m_2^2,q^2)&=&-2+\log\frac{m_1m_2}{\mu^2}+\frac{m_1^2-m_2^2}{q^2}\log\left(\frac{m_1}{m_2}\right)
	\nonumber\\&&
	+\frac{1}{q^2}\sqrt{|(m_1+m_2)^2-q^2||(m_1-m_2)^2-q^2|}f(m_1^2,m_2^2,q^2),
\end{eqnarray}
where
\begin{equation}
	f(m_1^2,m_2^2,q^2)=\left\{
\begin{array}{ll}
	\log\frac{\sqrt{(m_1+m_2)^2-q^2}-\sqrt{(m_1-m_2)^2-q^2}}{\sqrt{(m_1+m_2)^2-q^2}+\sqrt{(m_1-m_2)^2-q^2}}
	&\quad q^2\leq(m_1-m_2)^2\\
	2\arctan\sqrt{\frac{q^2-(m_1-m_2)^2}{(m_1+m_2)^2-q^2}}
	&\quad (m_1-m_2)^2<q^2<(m_1+m_2)^2\\
	\log\frac{\sqrt{q^2-(m_1-m_2)^2}+\sqrt{q^2-(m_1+m_2)^2}}{\sqrt{q^2-(m_1-m_2)^2}-\sqrt{q^2-(m_1+m_2)^2}}
	&\quad q^2\geq(m_1+m_2)^2\\
\end{array}\right.,
\end{equation}
and
\begin{small}
\begin{align}
	b_1(m_1^2,m_2^2,q^2)&=-\frac{1}{2}\left[\frac{m_1^2}{q^2}\left(\log\frac{m_1^2}{\mu^2}-1\right)-\frac{m_2^2}{q^2}\left(\log\frac{m_2^2}{\mu^2}-1\right)\right]+\frac{1}{2}\frac{m_1^2-m_2^2+q^2}{q^2}b_0(m_1,m_2,q)\\\nonumber
	b_2(m_1^2,m_2^2,q^2)&=\frac{1}{18}+\frac{1}{6}\left[\frac{m_1^2(2m_1^2-2m_2^2-q^2)}{(q^2)^2}\log\frac{m_1^2}{\mu^2}
	+\frac{m_2^2(2m_2^2-2m_1^2-q^2)}{(q^2)^2}\log\frac{m_2^2}{\mu^2}\right]\\\notag
    &-\frac{1}{3}\left(\frac{m_1^2-m_2^2}{q^2}\right)^2-\frac{1}{6}\left[2\left(\frac{m_1^2-m_2^2}{q^2}\right)^2-\left(\frac{m_1^2+m_2^2+q^2}{q^2}\right)\right]b_0(m_1,m_2,q).
\end{align}
\end{small}

\subsection{Some examples about tree and loop level contribution}
According to \cite{Zhang:2012cd}, we can calculate the loop contribution of following observables. 
\label{theory_tree}
\begin{enumerate}
    \item the decay width of $Z\to e^{+}e^{-}$, $\Gamma_{Z\to ee}$.
    numerical expressions:
\begin{align}
    \delta \Gamma^\mathrm{tree}_{Z\to ee}&=
    (10.7 c^{(1)}_{\varphi l}-1.15 c^{(3)}_{\varphi
    	l}-2.96 c_{\varphi D}-9.40 c_{\varphi e}-2.08
    c_{\varphi WB}+5.93 c^{'}_{ll})\times 10^{-3}.\\
        \delta\Gamma^\mathrm{loop}_{Z\to
            ee}&=
        (-124c^{(-)}_{\varphi Q}+3.24
        c^{(+)}_{\varphi Q}+140c_{\varphi t}
        -4.56c_{\varphi b}-2.11c_{\varphi tb}\\\notag
        &\quad+12.3 c_{tW}+1.33c_{bW}
        +17.6 c_{tB}+4.45c_{bB})\times10^{-6}.
       \end{align}
    \item the decay width of $W\to l\nu_{l}$, $\Gamma_{W\to l\nu_{l}}$.
    \begin{align}
    &\delta \Gamma^\mathrm{tree}_{W\to l\nu_{l}}=(-1.77c^{(3)}_{\varphi
    l}-1.45c_{\varphi D}-3.21c_{\varphi WB}+2.23c^{'}_{ll})\times 10^{-2}\\
    &\delta\Gamma^\mathrm{loop}_{W\to l\nu_{l}}=
	(-537c^{(-)}_{\varphi Q}+126c^{(+)}_{\varphi
		Q}+629c_{\varphi t}-34.3c_{\varphi b}
	-12.8c_{\varphi tb}\\\notag
    &\quad-64.7c_{tW}+8.53c_{bW}+245c_{tB}
	+41.5c_{bB})\times10^{-6}.
	\end{align}
    \item the coupling between the axial-vector current and $Z$
    boson and the coupling between the vector current
    and $Z$ boson in the process $\nu_{\mu}-e$ scattering
    process at low energy $g^{\nu_{\mu}e}_{LV}$, $g^{\nu_{\mu}e}_{LA}$ \cite{Erler:2004cx}. 
    \begin{small}
    \begin{align}  
        \delta g^{\nu_{\mu}e}_\mathrm{LVtree}&=   
        (-2.83c^{(1)}_{\varphi l}+5.32c^{(3)}_{\varphi
            l}+2.13c_{\varphi D}-3.03c_{\varphi e}\\
            \notag
            &+9.63 c_{\varphi WB}-3.03c_{le}-6.06c_{ll}-4.27c^{'}_{ll})\times10^{-2},\\
        \delta g^{\nu_{\mu}e}_\mathrm{LAtree}&=(1.51c_{\varphi D} + 3.03
        c_{\varphi e} + 3.03c_{le} - 6.06c_{ll} - 
        3.03c^{'}_{ll})\times10^{-2}. \\
        \delta g^{\nu_{\mu} e}_\mathrm{LVloop}&=
    (814c^{(-)}_{\varphi Q}+6.73
    c^{(+)}_{\varphi Q}-896c_{\varphi t}+213c_{\varphi
        b}
    +15.2c_{\varphi tb}\\\notag
    &\quad-715c_{t W}+111
    c_{b W}-747c_{t B}-170c_{bB})\times10^{-6},\\
    \delta g^{\nu_{\mu} e}_\mathrm{LAloop}&=(647c^{(-)}_{\varphi
        Q}+87.1c^{(+)}_{\varphi Q}-729c_{\varphi t}-5.5072c_{\varphi b}+10.8c_{\varphi tb})\times10^{-6}.
	    \end{align}    
\end{small}
\end{enumerate}

\subsection{\texorpdfstring{$e^{+}e^{-}\to W^{+}W^{-}$ and $e^{+}e^{-}\to W^{+}W^{-}\to l\nu_{l}ud$}{ee->WW and ee->WW->lvlud}}
The $W$ pair production cross-section and $W^{-}$ angular distribution are measured at LEP-II. However, MadGraph can not calculate the change of the space phase induced by $\delta m^{2}_{W}(m^{2}_{W})$, $\delta m_{W}$, and $\delta\Gamma_{W}$. $\delta\Gamma_{W}$ is the third-generation quark loop and tree level contribution to the decay width of $W$, $\Gamma_{W}$ respectively. Therefore, calculation of the effect induced by $\delta m^{2}_{W}(m^{2}_{W})$, $\delta m_{W}$, $\delta\Gamma_{W}$ is done with Feyncalc \cite{Shtabovenko:2016olh}\cite{Mertig:1990an} and Feynarts \cite{Kublbeck:1990xc}. The rest of the contribution from dimension-6 operators to the $W$ pair production cross-section and $W^{-}$ angular distribution can be calculated with MadGraph. Next, we demonstrate how to calculate these contributions with Feynarts and Feyncalc in details.
\subparagraph{$\delta\sigma_{\rm{on-shell}}$}: $\delta\sigma_{\rm{on-shell}}$ is the contribution of $\delta m^{2}_{W}(m^{2}_{W})$ and $\delta m_{W}$ to $\sigma_{\rm{on-shell}}$. $\sigma_{\rm{on-shell}}$ is the $W$ pair production cross-section in the process of $e^{+}e^{-}\to W^{+}W^{-}$ at SM tree level. 
\begin{equation}
    \delta\sigma_{\rm{on-shell}}=\partial_{m_{W}}\sigma_{\rm{on-shell}}\Big(\frac{\delta m^{2}_{W}(m^{2}_{W})}{2m_{W}}+\delta m_{W}\Big).
\end{equation}

\subparagraph{$\frac{d^{i} \delta\sigma_{\rm{off-shell}}}{d \cos\theta}$}: $\frac{d^{i} \delta\sigma_{\rm{off-shell}}}{d \cos\theta}$ is the contribution of $\delta m^{2}_{W}(m^{2}_{W})$, $\delta m_{W}$, $\delta\Gamma_{W}$ to $\frac{d^{i}\sigma_{\rm{off-shell}}}{d\cos\theta}$. $\frac{d^{i}\sigma_{\rm{off-shell}}}{d\cos\theta}$ is the $W^{-}$ angular distribution ($\theta$ is the polar angle between the $W^{-}$ and the $e^{-}$ beam) in $i$th bin at SM tree level in the process of $e^{+}e^{-}\to W^{+}W^{-}\to l\nu_{l}ud$. 

We can use of the phase space recursion relation and the narrow width approximation \cite{Han:2005mu} to calculate $\frac{d^{i}\sigma_{\rm_{off-shell}}}{d \cos\theta}$(this approximation will have only 1\% deviation from total cross-section).

\begin{equation}
	\frac{d^{i}\sigma_{\rm{off-shell}}}{d\cos\theta}\approx \frac{d^{i}\sigma_{\rm{on-shell}}}{d\cos\theta}
	\frac{\Gamma_{W\to l\nu_{l}}}{\Gamma_{W}}\frac{\Gamma_{W\to ud}}{\Gamma_{W}}
\end{equation}
where $\Gamma_{W\to ud}$ is the decay width of $W\to ud$ at SM tree level, $\frac{d^{i}\sigma_{\rm{on-shell}}}{d\cos\theta}$ is the $W^{-}$ angular distribution in $i$th bin at SM tree level in the process of $e^{+}e^{-}\to W^{+}W^{-}$. Then $\frac{d^{i} \delta\sigma_{\rm{off-shell}}}{d\cos\theta}$ is
\begin{align}
	\frac{d^{i} \delta\sigma_{\rm{off-shell}}}{d\cos\theta}&=\frac{d^{i} \delta\sigma_{\rm{on-shell}}}{d \cos\theta}\frac{\Gamma_{W\to l\nu_{l}}}{\Gamma_{W}}\frac{\Gamma_{W\to ud}}{\Gamma_{W}}+\frac{d\sigma_{\rm{on-shell}}}{d\cos\theta} \frac{\delta\Gamma_{W\to l\nu_{l}}}{\Gamma_{W}}\frac{\Gamma_{W\to ud}}{\Gamma_{W}}+\\
	&\frac{d\sigma_{\rm{on-shell}}}{d\cos\theta} \frac{\Gamma_{W\to l\nu_{l}}}{\Gamma_{W}}\frac{\delta\Gamma_{W\to ud}}{\Gamma_{W}}-2\frac{d\sigma_{\rm{on-shell}}}{d\cos\theta}
	\frac{\Gamma_{W\to l\nu_{l}}}{\Gamma_{W}}\frac{\Gamma_{W\to ud}}{\Gamma_{W}}\frac{\delta\Gamma_{W}}{\Gamma_{W}}
\end{align}
where $\delta\Gamma_{W\to l\nu_{l}}$, and $\delta\Gamma_{W\to ud}$ are the $\delta m^{2}_{W}(m^{2}_{W})$ and $\delta m_{W}$ contribution to $\Gamma_{W\to l\nu_{l}}$, $\Gamma_{W\to ud}$ respectively. $\frac{d^{i} \delta\sigma_{\rm{on-shell}}}{d \cos\theta}$ is the $\delta m^{2}_{W}(m^{2}_{W})$ and $\delta m_{W}$ contribution to $\frac{d^{i}\sigma_{\rm_{off-shell}}}{d \cos\theta}$. $\frac{d^{i}\sigma_{\rm_{off-shell}}}{d \cos\theta}$ is $W^{-}$ angular distribution in $i$th bin in the process of $e^{+}e^{-}\to W^{+}W^{-}$ at SM tree level.
	
However, the complete $\mathcal{O}(\alpha)$ electroweak (EW) corrections can not be ignored. We need calculate contribution of dimension-6 operators including the complete $\mathcal{O}(\alpha)$ electroweak (EW) corrections.

\subsubsection{\texorpdfstring{The contribution of $\delta m^{2}_{W}(m^{2}_{W}), \delta m_{W}$ to $\sigma_{\rm{on-shell}}^\mathrm{NLO}$}{The contribution of delta m2W(m2W), delta mW to sigmaonshellNLO}}
In the process of $e^{+}e^{-}\to W^{+}W^{-}$, our observables are $\sigma_{\rm{on-shell}}^\mathrm{NLO}$  at different $\sqrt{s}$. $\sigma_{\rm{on-shell}}^\mathrm{NLO}$ is the total cross-section of $W$ pair production including complete $\mathcal{O}(\alpha)$ electroweak (EW) correction in the process of $e^{+}e^{-}\to W^{+}W^{-}$. The assumption is made that the operators' contribution to $\sigma_{\rm{on-shell}}^\mathrm{NLO}$ can be calculated in this way:
\begin{equation}
    \delta\sigma_{\rm{on-shell}}^\mathrm{NLO}\approx\frac{\sigma_{\rm{on-shell}}^\mathrm{NLO}}{\sigma_{\rm{on-shell}}}\delta\sigma_{\rm{on-shell}}
\end{equation}
where  $\delta\sigma_{\rm{on-shell}}^\mathrm{NLO}$ is the contribution of $\delta m^{2}_{W}(m^{2}_{W})$ and $\delta m_{W}$ to $\sigma_{\rm{on-shell}}^\mathrm{NLO}$.

\subsubsection{\texorpdfstring{The contribution of $\delta m^{2}_{W}(m^{2}_{W})$, $\delta m_{W}$, $\delta\Gamma_{W}$  to $\frac{d^{i}\sigma_{\rm_{off-shell}}^\mathrm{NLO}}{d \cos\theta}$}{The contribution of delta m2W, delta mW, deltaGammaW to diff-xsec}}
In the process of $e^{+}e^{-}\to W^{+}W^{-}\to l\nu_{l}ud$, the observables are $\frac{d^{i}\sigma_{\rm_{off-shell}}^\mathrm{NLO}}{d\cos\theta}$. $\frac{d^{i}\sigma_{\rm_{off-shell}}^\mathrm{NLO}}{d\cos\theta}$ is $W^{-}$ angular distribution in the $i$th bin including the complete $\mathcal{O}(\alpha)$ electroweak(EW) correction. Since $\frac{d^{i}\sigma_{\rm_{off-shell}}^\mathrm{NLO}}{d\cos\theta}$ change very slowly in every certain bin($\cos\theta_{i}=(i-1)\times0.2$), it can be regarded as $\frac{d\sigma_{\rm_{off-shell}}^\mathrm{NLO}}{d\cos\theta}\arrowvert_{\cos\theta=\frac{\cos\theta_{i}+\cos\theta_{i+1}}{2}}$.   
The method of reweighting is applied to calculate $\frac{d^{i}\delta\sigma_{\rm_{off-shell}}^\mathrm{NLO}}{d \cos\theta}$ from $ \frac{d^{i}\delta\sigma_{\rm{off-shell}}}{d\cos\theta}$. $\frac{d^{i}\delta\sigma_{\rm{off-shell}}}{d\cos\theta}$ and $ \frac{d^{i}\delta\sigma_{\rm_{off-shell}}^\mathrm{NLO}}{d \cos\theta}$ are the contribution of $\delta m^{2}_{W}(m^{2}_{W})$, $\delta m_{W}$, $\delta\Gamma_{W}$ to $\frac{d^{i}\sigma_{\rm_{off-shell}}}{d \cos\theta}$ and $ \frac{d^{i}\sigma_{\rm_{off-shell}}^\mathrm{NLO}}{d \cos\theta}$ respectively
\begin{equation}
	\frac{d^{i}\delta\sigma_{\rm_{off-shell}}^\mathrm{NLO}}{d \cos\theta}\approx\frac{\frac{d^{i}\sigma_{\rm_{off-shell}}^\mathrm{NLO}}{d\cos\theta}}{\frac{d^{i}\sigma_{\rm{off-shell}}}{d\cos\theta}}\frac{d^{i}\delta\sigma_{\rm{off-shell}}}{d\cos\theta},
\end{equation}
The numerical result of $ d^{1}\delta\sigma_{\rm_{off-shell}}^\mathrm{NLO}$ and  when $\sqrt{s}$ is 182.66GeV:
\begin{footnotesize}
\begin{align}
	d^{1}\delta\sigma_{\rm_{off-shell}}^\mathrm{NLO}=&(-114.082c^{(-)}_{\varphi Q}+27.4928 c^{(+)}_{\varphi Q} + 132.446c_{\varphi t} - 3615.79c^{(3)}_{\varphi l}  \\\notag
	&-361.279c^{(3)}_{\varphi q}-7.22494c_{\varphi b} - 2.66921c_{\varphi tb} - 5.95838c_{t W}\\\notag
	&+2.23408c_{b W}+ 51.7414c_{tB} + 8.73857c_{bB} - 3260.26c_{\varphi D} \\\notag
	&-7201.74c_{\varphi WB}+ 1446.62c^{'}_{ll})\times 10^{-5}.
\end{align}	
\end{footnotesize}





\bibliography{ewfit}



\end{document}